\documentclass[fleqn,usenatbib]{mnras}

\usepackage{newtxtext,newtxmath}
\usepackage[T1]{fontenc}
\usepackage{ae,aecompl}
\usepackage{graphicx}
\usepackage{amsmath}
\usepackage{todonotes}
\usepackage{xspace}

\newcommand{\msun}{\,{M$_{\odot}$}\xspace}
\newcommand{\smassdflt}{$8.5 \leq \log_{10} \left(M_\star/\rm{M}_\odot\right) \leq 9.5$\xspace}
\newcommand{\smassdflthigh}{$10.5 \leq \log_{10} \left(M_\star/\rm{M}_\odot\right) \leq 11.5$\xspace}
\newcommand{\smassdfltlow}{$8.5 \leq \log_{10} \left(M_\star/\rm{M}_\odot\right) \leq 9.0$\xspace}
\newcommand{\smassmini}{$8.45 \leq \log_{10} \left(M_\star/\rm{M}_\odot\right) \leq 8.55$\xspace}
\newcommand{\hmassmini}{$10.6 \leq \log_{10} \left(M_h/\rm{M}_\odot\right) \leq 10.8$\xspace}
\newcommand{\Lya}{\ifmmode{\mathrm{Ly}\alpha}\else Ly$\alpha$\xspace\fi}
\newcommand{\angstrom}{\text{\normalfont\AA}}

\title[Lyman-$\alpha$ halos in TNG50]{The physical origins and dominant emission mechanisms \\of Lyman-alpha halos: results from the TNG50 simulation in comparison to MUSE observations}

\author[C. Byrohl et al]{Chris Byrohl$^{1}$\thanks{E-mail: cbyrohl@mpa-garching-mpg.de}, 
Dylan Nelson$^{1,2}$,
Christoph Behrens$^{3}$,
Ivan Kostyuk$^{1}$, 
\newauthor
Martin Glatzle$^{1}$, 
Annalisa Pillepich$^{4}$,
Lars Hernquist$^{5}$,
Federico Marinacci$^{6}$,
\newauthor
Mark Vogelsberger$^{7}$
\\\\
$^{1}$Max-Planck-Institut f\"{u}r Astrophysik, Karl-Schwarzschild-Str. 1, 85748 Garching, Germany \\
$^{2}$Universit\"{a}t Heidelberg, Zentrum f\"{u}r Astronomie, Institut f\"{u}r theoretische Astrophysik, Albert-Ueberle-Str. 2, 69120 Heidelberg, Germany \\
$^{3}$Institut f\"{u}r Astrophysik, Georg-August Universit\"{a}t G\"{o}ttingen, Friedrich-Hund-Platz 1, 37075 G\"{o}ttingen, Germany \\
$^{4}$Max-Planck-Institut f\"{u}r Astronomie, K\"{o}nigstuhl 17, 69117 Heidelberg, Germany \\
$^{5}$Harvard-Smithsonian Center for Astrophysics, 60 Garden Street, Cambridge, MA, 02138, USA \\
$^{6}$Department of Physics \& Astronomy, University of Bologna, via Gobetti 93/2, 40129 Bologna, Italy \\
$^{7}$Department of Physics, Massachusetts Institute of Technology, Cambridge, MA 02139, USA
}

\date{}
\pubyear{2020}

\begin{document}
\label{firstpage}
\pagerange{\pageref{firstpage}--\pageref{lastpage}}
\maketitle

\begin{abstract}
Extended Lyman-alpha emission is now commonly detected around high redshift galaxies through stacking and even on individual basis. Despite recent observational advances, the physical origin of these Lyman-alpha halos (LAHs), as well as their relationships to galaxies, quasars, circumgalactic gas, and other environmental factors remains unclear. We present results from our new Lyman-alpha full radiative transfer code \textsc{voroILTIS} which runs directly on the unstructured Voronoi tessellation of cosmological hydrodynamical simulations. We make use of the TNG50 simulation and simulate LAHs from redshift $z=2$ to $z=5$, focusing on star-forming galaxies with $8.0 < \log_{10}{(M_\star/\rm{M}_\odot)} < 10.5$. While TNG50 does not directly follow ionizing radiation, it includes an on-the-fly treatment for active galactic nuclei and ultraviolet background radiation with self-shielding, which are important processes impacting the cooling and ionization of the gas. Based on this model, we present the predictions for the stacked radial surface brightness profiles of Ly$\alpha$ as a function of galaxy mass and redshift. Comparison with data from the MUSE UDF at $z>3$ reveals a promising level of agreement. We measure the correlations of LAH size and central brightness with galaxy properties, finding that at the masses of $8.5 \leq \log_{10} \left(M_\star/\rm{M}_\odot\right) \leq 9.5$, physical LAH sizes roughly double from $z=2$ to $z=5$. Finally, we decompose the profiles into contributions from diffuse emission and scattered photons from star-forming regions. In our simulations, we find rescattered photons from star-forming regions to be the major source in observed LAHs. Unexpectedly, we find that the flattening of LAH profiles at large radii becomes dominated by photons originating from other nearby halos rather than diffuse emission itself.
\end{abstract}


\section{Introduction}

The Lyman-$\alpha$ (\Lya) line of hydrogen at $121.567$\,nm is one of the brightest emission lines in the Universe. It allows us to detect, and trace the distribution of, galaxies even out to very high redshifts $z>5$. These Lyman-$\alpha$ emitters \citep[LAEs;][]{Partridge1967} can be used to probe the physics of galaxy formation \citep{finkelstein09,nagamine10,erb14} as well as constrain cosmological parameters and large-scale structure \citep{Hill2008,Adams2011}. Starting in the 80s, spatially extended \Lya emission has been detected, often called Lyman-$\alpha$ blobs (LABs) and Lyman-$\alpha$ nebulae with large extents between $\sim 10-100\ $pkpc~\citep{McCarthy1987,heckman91,Steidel2000}.

More recently, Lyman-$\alpha$ halos (LAHs) have been discovered around star-forming galaxies that show \Lya emission far beyond the galaxies' optical bodies, tracing the circumgalactic rather than interstellar gas~\citep[e.g.][]{hayes13}. 
While LAHs are fainter and smaller than LABs in their \Lya extent, they might be a generic feature around \Lya emitting galaxies. 
Narrow-band imaging can efficiently detect LAHs at targeted redshifts through stacking \citep{hayashino04,Steidel2011,Matsuda2012,Feldmeier2013}, and narrow-band surveys enable ultra-deep, blind samples of LAHs around distant galaxies \citep{Momose2014,Momose2016,Kakuma2019}.
In the last years modern surveys performed with integral field unit (IFU) spectrographs on 10m-class telescopes, such as the Multi Unit Spectroscopic Explorer (MUSE) and the Keck Cosmic Web Imager (KCWI), take place. These new instruments allow the study of individual, faint LAHs opposed to previous narrow-band stacks. Along with the IFUs' spectral resolution these recent surveys largely increase the information available from LAH observations. 
Hundreds of individually extended Lyman-$\alpha$ halos at $z\gtrsim 2$ have been revealed since\citep{Wisotzki2016}. Many of these are specifically targeted samples which focus on bright quasars \citep{borisova16,Sullivan20,Cai2019,guo20}, based on strong earlier evidence of enhanced Ly$\alpha$ emission around active galactic nuclei (AGN) \citep{cantalupo14,battaia16,farina19,battaia19}. Others exploit the large field of view of MUSE, in particular, to conduct blind surveys for LAHs around more typical, generally star-forming galaxies \citep{leclercq17,wisotzki18,leclercq20}. At the same time, follow-up with other instruments such as ALMA reveals complementary views on other gas phases within LAHs including CO \citep{emonts19}.

Beyond the circumgalactic medium (CGM), first attempts have been made to detect cosmic web filaments in \Lya emission \citep{Gallego2018,umehata19,lusso19}, as well as to study the interface between the IGM and CGM as gas flows feed galaxies \citep{Martin2019}. Finally, Ly$\alpha$ emission is also a powerful cosmological tool. The Hobby-Eberly Telescope Dark Energy Experiment (HETDEX) is an IFU survey which will detect up to a million LAEs, as well as many extended LAHs, at more moderate spatial and spectral resolution \citep{Hill2008}. 

Despite extensive observational detection and characterization, the physical properties and nature of Ly$\alpha$ halos remains an open topic. It is unclear whether the extended profiles are mainly sourced by (i) diffuse emission outside of central galaxies, or whether (ii) scatterings of Ly$\alpha$ photons emitted from within central galaxies power observed LAHs. Diffuse emission is commonly considered to be sourced by gravitational cooling \citep{Haiman2000,Fardal2001,Faucher2010} and fluorescence \citep{Gould1996,Cantalupo2005,Kollmeier2010,MasRibas2016}, while star-formation and quasars can provide significant emission within galaxies \citep{Dijkstra2006,Zheng2011b} that can scatter with neutral hydrogen in the CGM. Emission from orbiting satellite galaxies can also lead to extended \Lya profiles \citep{MasRibas2017}.

The difficulties in determining the powering source of LAHs are closely linked to the resonant nature of \Lya photons that can scatter many times in astrophysical environments before escaping towards the observer. This causes the observed frequency and angular position to significantly change due to radiative transfer (RT). Only in the simple, symmetric geometries, RT can be solved analytically~\citep[][]{Harrington1973,Neufeld1990,Loeb1999,Lao2020}. Moving to more realistic setups and in particular hydrodynamical simulations, RT has to be solved numerically.

Recently our theoretical understanding of the \Lya emission around galaxies has been pushed forward with the development of cosmological hydrodynamical simulations~\citep{Vogelsberger2020} which are able to produce broadly realistic galaxy populations \citep{Vogelsberger2014a,Vogelsberger2014b,Genel2014,schaye15,dubois16,pillepich18b,dave19}. Crucially, these simulations predict the full distribution of gas, including neutral hydrogen, in and around galaxies. With volumes of many Mpc a side, these simulations include thousands of well resolved galaxies across the mass spectrum. This enables explicit \Lya RT calculations to solve for the propagation and scattering of Ly$\alpha$ photons through the interstellar, circumgalactic, and intergalactic media \citep{Laursen2009,Behrens2018}. Due to computational expense, \Lya RT on these volumes can for now exclusively be done in post-processing, even though \Lya pressure can be dynamically important~\citep{Smith2019}. 
Similarly, a proper treatment of the temperature and ionization state of the gas requires hydrodynamics to be coupled with the radiative transfer of Lyman Continuum (LyC) photons. While these radiation-hydrodynamical (RHD) simulations are becoming increasingly feasible, this remains the case only for individual zoom simulations~\citep[e.g.][]{Rosdahl2012,mitchell21} and for cosmological volumes only down to high redshifts around $z\sim 5$~\citep{Gnedin2017,Rosdahl2018,Ocvirk2020}.

There have been various recent efforts to understand the Ly$\alpha$ halos in emission by post-processing hydrodynamical simulations \citep[e.g.][]{Lake2015,Smith2019,Kimock2020} with \Lya RT. These works typically consider one, to a few tens, of galaxies -- rather than full cosmological volumes -- making it difficult to draw conclusions about the environmental dependencies of LAHs. Notable exceptions are \cite{Gronke2017}, who predicted LAH properties from the Illustris simulation, and \citet{Zheng2011b} who predicted low surface brightness Ly$\alpha$ emission from a cosmological reionization simulation ($z=5.7$), although with a limited, `halo-scale' hydrodynamical resolution $\sim 30\ $pkpc.

In these theoretical works the most important emission origin(s) and source(s) remain disputed. For example, \citet{Lake2015} find good agreement for their set of 9 LAHs with mass $10^{11.5}\ $ M$_\odot$ contrasted with data from \citet{Momose2014}, stressing the importance of gravitational cooling in the outer halo to explain the observed profiles, while \citet{Gronke2017} simulate Lyman-$\alpha$ nebulae with masses $10^{11.5}-10^{13.5}\ $ M$_\odot$ and find the simulation can produce \Lya extents as large and luminous as those observed, only using central emission from AGN and star-formation. 

Recently, cosmological volumes from modern hydrodynamical simulations have been studied in the context of \Lya emission focusing on the detectability of the cosmic web~\citep{Elias2020,Witstok21}. Latest observations of the \Lya cosmic web in~\cite{Bacon2021} might point at the importance of emission from (faint) galaxies that has been missing in former theoretical explorations.

In this work we improve on several aspects of previous computational studies, revisiting the nature of Lyman-$\alpha$ halos. Specifically, we couple the new, high-resolution cosmological magnetohydrodynamical simulation TNG50 \citep{pillepich19,nelson19b} of the IllustrisTNG project to our new radiative transfer code \textsc{voroILTIS}. The former provides a competitive combination of volume (a statistically robust sample of $\sim 6,800$ galaxies with $M_\star > 10^7$\msun at $z=2$) and resolution ($\sim$ 100 parcsecs in the ISM, albeit with a simplified sub-grid treatment of the cold phase, and $<1$\,kpc in the CGM). The latter includes several Ly$\alpha$ emission models and a Monte Carlo treatment of the scattering process directly on the full Voronoi tessellation of the gas distribution of the entire simulation volume, enabling a self-consistent treatment of IGM attenuation \citep{byrohl20a}. We note that the TNG simulations are not RHD, and omit on-the-fly radiation transport.
Our setup enables us to statistically contrast the simulation predictions to existing LAH observations, while also probing questions regarding the dominant origins, emission sources and relevance of rescattering for the existence of LAHs, and making future predictions in as of yet unobserved regimes.

The structure of this paper is as follows. In Section~\ref{sec:methods}, we describe our radiative transfer code \textsc{voroILTIS}, the Ly$\alpha$ emission model, and the analysis details of the underlying IllustrisTNG simulations on which the radiative transfer code is run. In Section~\ref{sec:results} we present the results for the radial profiles and related reduced quantities from our simulations and a comparison to observational data. In Section~\ref{sec:discussion} we discuss the radial profile shapes and reduced quantities in more detail. We summarize our findings in Section~\ref{sec:conclusions}.
We also point to an extensive Appendix~\ref{sec:modelassumptions} that highlights different modeling assumptions in the underlying TNG50 simulations and the \Lya radiative transfer.

\section{Methods}
\label{sec:methods}

\subsection{IllustrisTNG and TNG50}
\label{sec:sims}

We use the outcome of the IllustrisTNG simulations -- both the galaxy properties as well as gas distributions -- as the basis for our radiative transfer simulations of Lyman-$\alpha$ halos. The IllustrisTNG simulations \citep[hereafter, TNG;][]{pillepich18b,naiman18,nelson18a,marinacci18,springel18} are a series of three large-volume magnetohydrodynamical cosmological simulations of galaxy formation. All are run with the \textsc{AREPO} code \citep{Springel2010}, which solves the coupled equations of self-gravity and ideal, continuum magnetohydrodynamics \citep{pakmor11} with a `moving mesh' discretization based on an unstructured Voronoi tessellation of space.

The TNG galaxy formation model \citep{weinberger17,pillepich18a} is an evolution of the original Illustris model \citep{vogelsberger13,torrey14} and includes a number of important changes. All TNG simulations include models for the physical processes most important for galaxy formation: primordial and metal-line gas cooling, heating from the metagalactic background radiation field, star formation above a density threshold, stellar population evolution and chemical enrichment following supernovae Ia, II, and AGB stars, and the seeding, merging, and growth via accretion of supermassive black holes (SMBHs). With respect to the earlier Illustris model, two critical updates have been made. First, the galactic-scale winds launched by stellar feedback have been revised \citep{pillepich18a}, which impacts the gas (and stellar) contents of low mass galaxies in both their ISM and CGM. Further, TNG includes a two-mode blackhole feedback operating in a thermal `quasar' state at high accretion rates and a kinetic `wind' state at low accretion rates. The latter is a new model for low-state SMBH feedback, in the form of a time stochastic, directionally variable, high-velocity kinetic wind \citep{weinberger17}.

The temperature and ionization state of hydrogen -- crucial to \Lya emission and scattering -- is computed within TNG incorporating primordial cooling following~\cite{Katz1996} with additional metal-line cooling from \textit{CLOUDY} cooling tables. 
Both metal and primordial cooling are further modified by the assumption of a uniform, time-varying UV background using the intensities given in~\cite{Faucher2009} for photoionization and photoheating. 

Given the substantial impact of AGN, photoionization and photoheating due to their ionizing flux is incorporated through an AGN radiative feedback channel~\citep[for details see][]{vogelsberger13,weinberger17,weinberger18}. Intrinsic AGN luminosities are assigned strictly proportionally to the black hole accretion rate above a certain accretion threshold. 
Furthermore, an obscuration factor derived from observational work in \cite{Hopkins2007} sets the escape of ionizing radiation acting upon the surrounding gas under the assumption of optically thin gas. When an AGN with radiative feedback is present in TNG50, the resulting radiation field dominates over the UVB for the studied $\lesssim 50\ $pkpc around galaxies.

We discuss the AGN radiation field in TNG along with its impact on \Lya emission and reprocessing of \Lya photons for the radial profiles in Appendix~\ref{sec:ionizingsources}.

The ionizing radiation from the UVB and AGN entering the primordial cooling and \textit{CLOUDY} metal-line cooling is attenuated by a self-shielding factor based on radiation transfer simulations in~\cite{Rahmati2013}. These contributions are included on-the-fly, and not in post-processing.

The TNG model has been shown to produce galaxy and circumgalactic medium properties in broad agreement with available observational constraints. The properties of the CGM have been compared in terms of their OVI column densities \citep{nelson18b}, gas contents \citep{pillepich18a,terrazas20}, x-ray properties \citep{barnes18,truong20}, outflows and dynamics \citep{nelson19b}, and MgII covering fractions \citep{nelson20a}, but this is the first study of Lyman-$\alpha$ halos in TNG.

Of the three IllustrisTNG simulations, TNG50, TNG100, and TNG300 \citep{nelson19a} we here exclusively use the highest resolution box TNG50 \citep{nelson19b,pillepich19} with a gas mass resolution of $m_{\rm bayron} = 8.5 \times 10^4$\msun and a dark matter mass resolution of $m_{\rm DM} = 4.5 \times 10^5$\msun which is $\sim$\,15 ($\sim$\,120) times higher than TNG100 (TNG300). The corresponding spatial resolution of TNG50 is of order $\sim 100\ $physical parsecs in the dense ISM, and this small-scale structure is useful given the strong resolution dependence of \Lya radiative transfer at lower resolutions as demonstrated in~\cite{Behrens2018}, although a more sophisticated model for the cold phase of the ISM would be needed to fully capture \Lya RT effects on small scales.

TNG adopts a set of cosmological parameters consistent with recent results by the Planck collaboration \citep{planck16}, namely $\Omega_{\Lambda,0}=0.6911$, $\Omega_{m,0}=0.3089$, $\Omega_{b,0}=0.0486$, $\sigma_8=0.8159$, $n_s=0.9667$ and $h=0.6774$.

\subsection{voroILTIS}

The \Lya radiative transfer is calculated in post-processing on the TNG simulation output with an updated version of \textsc{ILTIS}\footnote{The public version of the \textsc{ILTIS} code is currently available at \url{github.com/cbehren/Iltis}, where the Voronoi version will also be released in the future.}, a light-weight line emission transfer code as presented in \cite{Behrens2019}. \textsc{ILTIS} implements a Monte Carlo approach, spawning single-wavelength photon `packages' (representing a large number of actual photons) at emission sites, and following their scattering as they traverse the underlying gas distribution. For efficiency, at each scattering event we output the attenuated luminosity contributions along specified line-of-sights towards assumed observers \citep[the `peeling-off' algorithm;][]{Whitney2011}. 

We have developed a new version of the code, \textsc{voroILTIS}, which runs directly on the unstructured mesh of a Voronoi tessellation. As IllustrisTNG uses this geometry to represent the gas distribution during the simulation and for the evolution of hydrodynamical quantities, no intermediate interpolation or re-sampling steps are required, and the density field in the RT calculation is self-consistent with the simulation. The mesh is re-created in postprocessing with a parallelized wrapper to the Voronoi tessellation code \textit{voro++} \citep{voroplusplus}. We then spawn photons for each cell in the mesh according to the local emissivity, an update which makes simulating diffuse emission feasible. 

After emission, photons propagate through the simulation domain, and across the periodic boundaries of the box as appropriate. Given the amplitude of the Hubble flow (several $100\ $km/s/Mpc) in the simulated redshift range and typical gas velocities, photons are quickly shifted into the far wings of the line profile where they have negligible cross-section. Therefore, we do not need to construct light-cones and simply propagate all photons for the necessary propagation length $l$ of order $\sim 10$\,Mpc. Here, we chose $l = 28\ $cMpc/h. An upcoming methods paper will document the code improvements used in the present work (\textcolor{blue}{Behrens \& Byrohl, in prep}). A predecessor of \textsc{ILTIS} has also been used in \cite{Behrens2018} and \cite{Byrohl2019} where additional details are available.

\begin{figure*}
\centering
\includegraphics[width=1.0\linewidth]{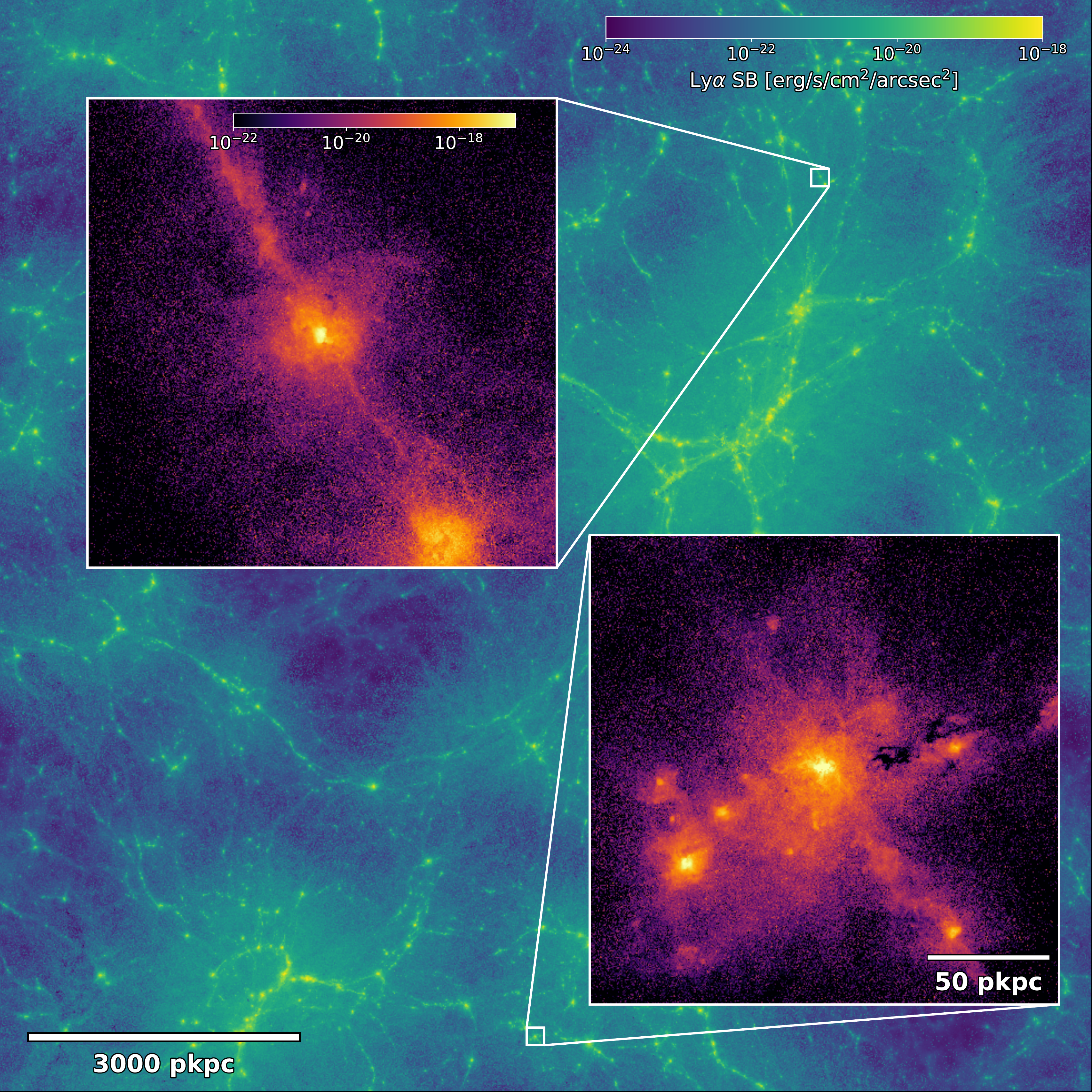}
\caption{\Lya surface brightness map for the entire TNG50 cosmological simulation at $z=3$, highlighting the large-scale structure of the cosmic web as seen in Ly$\alpha$ emission. We impose a Gaussian point spread function (PSF) with a FWHM of $0.7\ $arcsec at a binning resolution $\Delta_\mathrm{res}$ of $8.5\ $ckpc/h and project through a slice depth of $5.25$ cMpc/h. The inset panels (no PSF, $\Delta_\mathrm{res}=0.4\ $pkpc) show two individual Lyman-$\alpha$ halos, on the scale of the halo virial radii, for moderate mass objects: M$_\mathrm{200} \simeq 5\cdot 10^{10}$\msun and M$_\mathrm{200} \simeq 1.2\cdot 10^{11}$\msun (top and bottom, respectively). \Lya photons are predominantly emitted in the star-forming regions of the central galaxies, resonantly scatter and illuminate the more extended gaseous halos, including filamentary inflows. The more massive halo (lower right) has a number of star-forming satellites which also contribute Ly$\alpha$ emissivity and boost the local surface brightness.}
\label{fig:overviewplot}
\end{figure*}

\subsection{Lyman-alpha radiative processes}
\label{sec:emission}

\Lya photons are dominantly created by recombination of ionized hydrogen atoms with electrons, and de-excitation of excited neutral hydrogen. Different physical processes can power their creation: For recombination, ionizing radiation originates from star-forming regions as well as from the metagalactic ultraviolet background. For collisional de-excitation, the thermal state of the gas provides the gas heating mechanism. We refer to those processes to emission sources and in figures call them `rec', and `coll', respectively. 
As we discuss in the next paragraphs, we use a special description for the recombinations in star-forming regions, which we abbreviate as `SF' in figures.

We model recombinations after ionization by a spatially uniform background radiation field and AGN radiative feedback as implemented in TNG50. We assume case-B recombination, resulting in an emissivity

\begin{equation} \label{eq:lumrec}
\epsilon_\mathrm{rec} = f_\mathrm{rec}(T) \,n_\mathrm{e} \,n_\mathrm{HII} \,\alpha(T) \,E_{\mathrm{Ly}\alpha}
\end{equation}

\noindent where $n_\mathrm{e}$ and $n_\mathrm{HII}$ are the electron and ionized hydrogen number densities, $\alpha(T)$~\citep[from][]{Draine2011} is the recombination coefficient, $f_\mathrm{rec}$ is the recombination emission probability~\citep[from][]{Cantalupo2008} assuming case-B, and $E_{\mathrm{Ly}\alpha}$ is the \Lya transition energy. Note that $\alpha(T)$ and $f_{\rm rec}(T)$ are both temperature dependent.

In addition to recombinations, the de-excitation of excited hydrogen atoms can lead to the emission of Ly$\alpha$ sourced by collisional excitations depending on the gas thermal state. The rate is proportional to the colliding species, $n_\mathrm{e}$ and $n_\mathrm{HI}$. The luminosity density is then 

\begin{equation} \label{eq:lumexc}
\epsilon_\mathrm{coll} = \gamma_{\mathrm{1s2p}}(T) \,n_\mathrm{e} \,n_\mathrm{HI} \,E_{\mathrm{Ly}\alpha}.
\end{equation}

\noindent We take the collisional excitation coefficient $\gamma_\mathrm{1s2p}(T)$ from tabulated fits \citep{Scholz1990,Scholz1991}.

For star-forming gas cells, where the TNG model invokes a sub-grid effective equation of state model for the two-phase ISM \citep{Springel2003}, the simulation's temperature and hydrogen density do not reflect their physical meaning entering Eqn.~\eqref{eq:lumrec}. Hence, we instead derive the recombination rate from the star-formation rate. The ionization of hydrogen by young and massive stars followed by recombination dominates the local \Lya emission and the luminosity within galaxies is thus proportional to the star-formation rate. 

Here, we adopt a simple linear model for the Ly$\alpha$ luminosity density

\begin{equation} \label{eq:lumSF}
\epsilon_\mathrm{SF} = 
10^{42} \left(\frac{\dot{M}_\star}{\mathrm{M}_\odot \mathrm{yr}^{-1}}\right) \frac{\mathrm{erg/s}}{V_\star}
\end{equation}

\noindent where $\dot{M}_\star$ is the star-formation rate within a star-forming Voronoi cell's volume $V_\star$~\citep[see][]{Dijkstra2019}.

This relation is consistent with the H$\alpha$-SFR relation used in~\cite{Kennicutt1998} and an assumed \Lya to H$\alpha$ recombination ratio in the range of 8-10~\citep{Hummer1987}. It can be directly calculated by integrating the ionizing flux for a given stellar population but has large modeling uncertainties~\citep{Furlanetto2005}. 
Nevertheless this relation is commonly used in simulations and theory as an estimate for the \Lya emission from stellar populations~\citep[see e.g.][]{Furlanetto2005,Zheng2010,Behrens2018}. Equation~\eqref{eq:lumSF} does not account for \Lya destruction by dust in the complex subgrid ISM structure, and we do not take into consideration the stellar populations' age and metallicity. Incorporating these factors would add suppression and scatter in the relation; we discuss the modeling uncertainties and their implications for this paper in Section~\ref{sec:SFluminosity}. In the discussion, comparing to observations, we find that the modeling uncertainties are more important than the lack of dust attenuation for halos in the mass range \smassdflt{}. Nevertheless, a treatment of dust~\citep[see e.g. the effective treatment in][]{Lake2015,Inoue2018} and the introduction of sources of scatter in Equation~\eqref{eq:lumSF} are desirable for future work, and necessary at the high mass end.

The star-formation rate is taken directly from the TNG output in each cell volume. The TNG star-formation model is described in~\cite{pillepich18a}. In short, a gas cell is star-forming if and only if its physical density exceeds 0.1 hydrogen atoms per cubic centimeter, in which case collisionless star particles are stochastically formed~\citep[also see][]{Springel2003,vogelsberger13}. 

For recombination and de-excitations, we spawn one weighted photon package per Voronoi cell with the luminosity according to the models above. For star-formation, we only spawn photons in Voronoi cells bound to a halo of mass $M_{\rm 200,crit} > 10^{10}\ $M$_\odot$. In TNG50 this results in a sample of $\sim$\,5000 to $\sim$\,13000 LAHs at redshifts $z\in\left[2,5\right]$.

We found that a constant photon count per cell for Equations~\eqref{eq:lumrec} and~\eqref{eq:lumexc} results in good convergence for the radial profiles. In such a scheme, photon packages can carry vastly varying luminosity weights. This however is desired as we need to also trace faint, optically relatively thin regions in the outer CGM where surface brightnesses are orders of magnitudes lower than in the central regions of the LAHs.

In contrast, a Monte Carlo sampling relying on a fixed luminosity weight per photon package appears faster for the emission from star-forming cells (see Eqn.~\eqref{eq:lumSF}). For simplicity, we stick with the fixed Monte Carlo photon count per cell, but confirm that this photon count leads to a converged result (see Appendix~\ref{sec:photonconvergence}) for star-forming cells.

We refer to these photon packets as `photons' for brevity. Photons are always injected at the Ly$\alpha$ line-center in the rest-frame of the emitting gas cells. We have also explored different spectral emission distribution and generally find little impact -- details are given in Appendix \ref{sec:inputspectrum}.

In our model, we do not account for escaping ionizing radiation from star-forming regions that recombines outside of its emission region ("fluorescent radiation"). Fluorescence in our simulations hence only arises from the uniform UV background and the radiation field from AGN incorporated in TNG. In Appendix~\ref{sec:ionizingsources} we discuss the impact of ionizing sources by means of the AGN radiation implemented within the model underlying TNG50.
The radiative feedback from AGN is accounted for in that it affects the ionization state, temperature and cooling rates of the gas and hence can boost recombinations and collisional excitations. We do not consider additional recombinations in gas cells due to the activity of SMBHs.

\subsection{Post-processing and Observational Realism}
\label{sec:postprocessing}

The radiative transfer simulations provide us with large outputs of Monte Carlo photons. Each has a corresponding luminosity, frequency, emission source (see Section~\ref{sec:emission}), and positions of initial emission and last scattering. We save two distinct sets of photons:

\begin{itemize}
\item \textbf{``intrinsic''}: Ly$\alpha$ photons as directly emitted from gas cells, neglecting any subsequent interactions with gas.
\item \textbf{``processed''}: Peeling-off contributions after each scattering of propagated intrinsic Ly$\alpha$ photons. These photons include the modifications from scatterings, IGM attenuation and potential dust destruction.
\end{itemize}

Observed \Lya light corresponds to the `processed' photons only. Comparing results for intrinsic and processed photons allows us to quantify the redistribution of photons since their initial emission and the overall importance of \Lya RT in our astrophysical setting.

Both intrinsic and processed photons can be filtered based on the originating Voronoi cell, which is recorded by the intrinsic photons, and inherited by any peeling-off contribution. 

Using this, we can classify the emission origin of each photon according to four distinct categories:

\begin{itemize}
\item \textbf{``central galaxy''}: The photon originates in the central subhalo (i.e., galaxy) of the targeted halo.
\item \textbf{``outer halo''}: The photon originates within the targeted halo, but outside of the central galaxy.
\item \textbf{``IGM''}: The photon originates in a region not associated with any halo.
\item \textbf{``other halo''}: The photon originates from a region associated with a halo that is different from the targeted.
\end{itemize}
The definition of those categories relies on the halo and subhalo catalogues provided by IllustrisTNG, where halos and subhalos are defined via the Friends-of-Friends and Subfind algorithm respectively (see~\cite{nelson19a}). Each photon falls into exactly one of these categories.

First, we compute 2D surface brightness maps for all galaxies centered on the host halo position, using a pixel size of $0.8\ $pkpc. We include all scattered (i.e. processed) photons irrespective of their origin, as would be seen observationally.
As a result these maps include emission from diffuse gas around the halo and even emission from other nearby galaxies and halos. For the projection depth we include all scattered photons reaching the observer from within $\pm 100\ $pkpc around the galaxy along the line of sight. By adopting this simple prescription we effectively ignore \Lya frequency information (diffusion). Observational studies using integral field spectroscopy commonly adopt an adaptive wavelength window to incorporate all \Lya flux of the source based on their varying spectral widths. In our approach, where we can isolate frequency diffusion, we can forgo implementing such an adaptive algorithm. In the Appendix~\ref{sec:spectra}, we show the impact when incorporating spectral information and adopting a fixed wavelength window around each emitter. Our quantitative results on LAH sizes and qualitative behaviour at large radii remains unchanged irrespective of the chosen method of depth integration.

Unless otherwise stated, we always apply a Gaussian point spread function (PSF) with a FWHM of $0.7\ $arcsec. This PSF roughly corresponds to that of MUSE UDF data we compare in the later part of the paper (see Section \ref{sec:MUSEcomparison}).

\begin{figure*}
\centering
\includegraphics[width=1.0\linewidth]{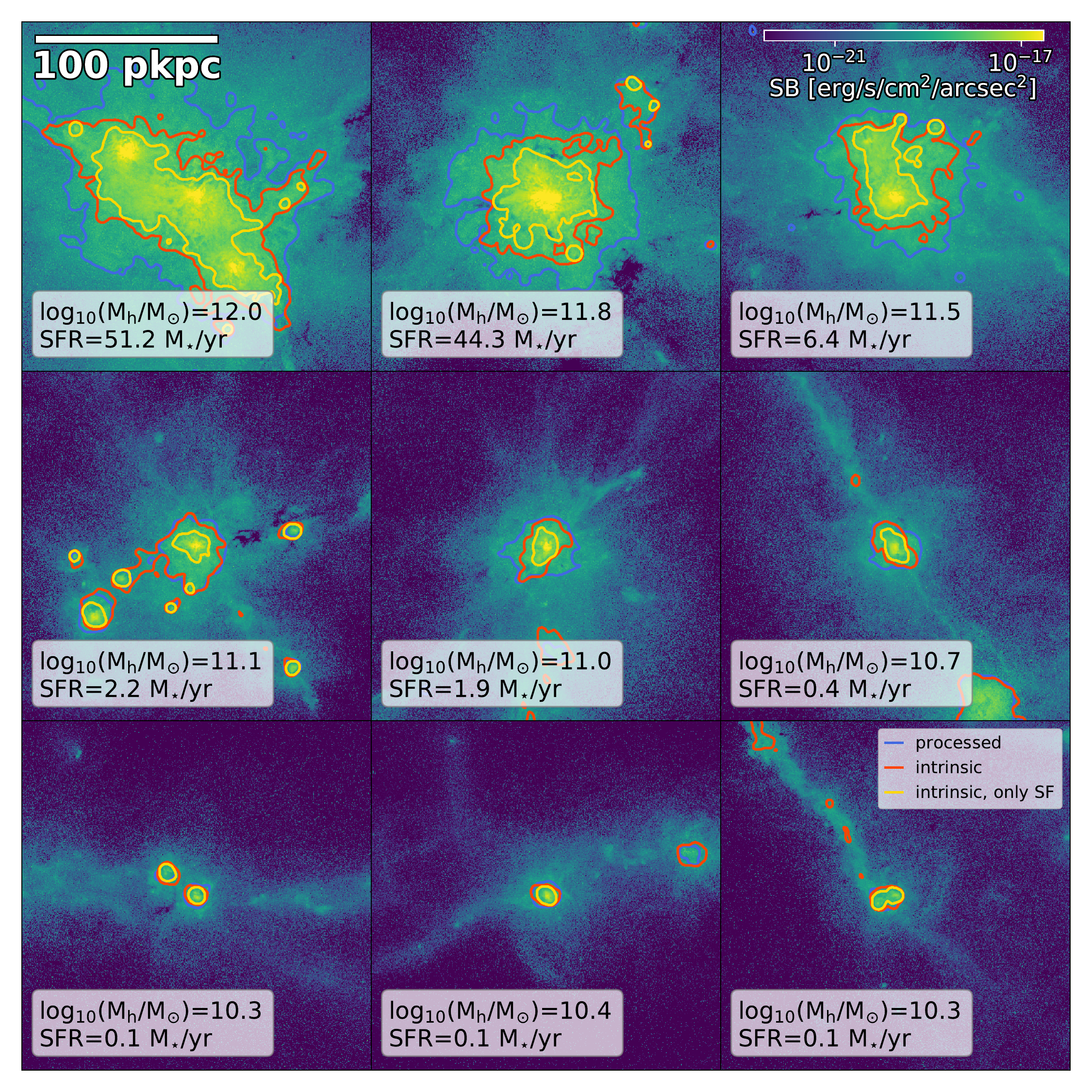}
\caption{Two-dimensional \Lya surface brightness maps of nine Lyman-$\alpha$ halos at $z=3.0$ in TNG50. The LAHs are ordered by their star-formation rate in a mass range of roughly $10^{10}$-$10^{12}\ $M$_\odot$. Contours highlight the surface brightness value of $10^{-19}\ $erg/s/cm$^{2}$/arcsec$^{2}$, showing the final observable result (blue; i.e. accounting for scattering and IGM attenuation) contrasted against intrinsic photons (red) and intrinsic photons due only to star-formation (yellow). We commonly find spatially extended intrinsic emission from star-formation in the most massive halos ($\log_{10}\left(M_{h}/M_{\odot}\right)\geq 11.5$), while intrinsic emission from other sources is even more extended. The scattering of Ly$\alpha$ photons expands the extent of high surface brightness features, particularly in the more massive halos. Here we adopt a pixel size of $\Delta_\mathrm{res}=0.4\ $pkpc (no PSF).}
\label{fig:panel3x3_total_z3_2Dviews}
\end{figure*}

\subsection{LAH Sample and Reduced Statistics}
\label{sec:reducedstats}

In our analysis we focus on halos with galaxy stellar masses of $8.0 < \log{(M_\star/\rm{M}_\odot)} < 10.5$. 
For those halos, one dimensional radial profiles are computed by averaging the pixel values for a given radial bin. We characterize these radial LAH profiles with a number of `reduced statistics': two measurements of Lyman-$\alpha$ halo size, the half-light radius $r_{1/2}$ and the exponential scale length $r_0$, and the `central' surface brightness value $\rm{SB}_0$, which we take as the value of the surface brightness map pixel(s) closest to the halo center, after smoothing by the PSF. 

The \Lya half-light radius $r_{1/2}$ is computed from the one-dimensional radial profile as the radius enclosing half of the total surface brightness contained within 50 pkpc. Because SB($r$) does not vanish at large distances due to contributions from other halos and diffuse gas, this measure depends on this chosen outer radius.

We also fit the one-dimensional radial profile with a single exponential $\rm{SB}(r) = \rm{SB}_{0,\mathrm{fit}}\exp(-r/r_0)$ \citep[as in][]{Cai2019}. The two parameters are the normalization $\rm{SB}_{0,\mathrm{fit}}$ and the scale length $r_0$. We fit the simulated profiles between 0.4 and 2 arcsec, taking a finite lower limit to exclude the impact of the PSF \citep[as is common observationally;][]{Wu2020} and the upper limit where the profile commonly transitions from an exponential to more flattened shape. Some observational studies fit a sum of two exponentials \citep{leclercq17,Wu2020}, but this method sets one exponential scale length to that of the UV light, adding additional modeling uncertainties which we avoid with the simpler size measure. For comparison with observations, we impose a simple signal-to-noise criterion, considering only data points with $S/N\geq 5$. We derive the noise from a Gaussian standard deviation of $2 \times 10^{-19}\ $erg/s/cm$^2$/arcsec$^2$ per pixel of the 2D SB maps. While definitions of LAH sizes vary, we impose the same fitting routine to all compared simulated and observed radial profiles for a fair comparison. Due to the variety of radial shapes for individual LAH, the exponential function can be a bad fit at times. We exclude such cases (<10\%) by imposing a maximum relative error of 10\% for either fit parameter as given from the estimated covariance matrix of the least square fit. We commonly specify the scale length of a given sample in the form of $\mathrm{median}^\mathrm{high}_\mathrm{low}$ where "low" and "high" are the 16th and 84th percentile.

In this work we refer to comoving units with a preceding `c', while physical units are explicitly preceded by `p'. We always present surface brightness in units of erg/s/cm$^2$/arcsec$^2$.

\section{Results}
\label{sec:results}

We begin in Figure \ref{fig:overviewplot} with a visual overview of \Lya emission from a large cosmological region encompassing thousands of individual emitters. Here we show the large-scale structure of the entire TNG50 simulation at $z=3$, with scattered \Lya photons illuminating not only LAHs, but also the cosmic web within which they reside. The zoom-in panels show two individual LAHs and their substructure (i.e. satellite galaxies), on the scale of the halo virial radius, for moderate halo masses of $M_\mathrm{200} \simeq 5 \times 10^{10}$\msun (top) and $M_\mathrm{200} \simeq 1.2 \times 10^{11}$\msun (bottom), where Ly$\alpha$ photons are predominantly emitted at the sites of star-formation in the central galaxy. These photons then resonantly scatter to illuminate extended gaseous halos of the circumgalactic medium (CGM), where the complex dynamics of high-redshift inflows mix with feedback-driven outflows.

\begin{figure*}
\centering
\includegraphics[width=1.0\linewidth]{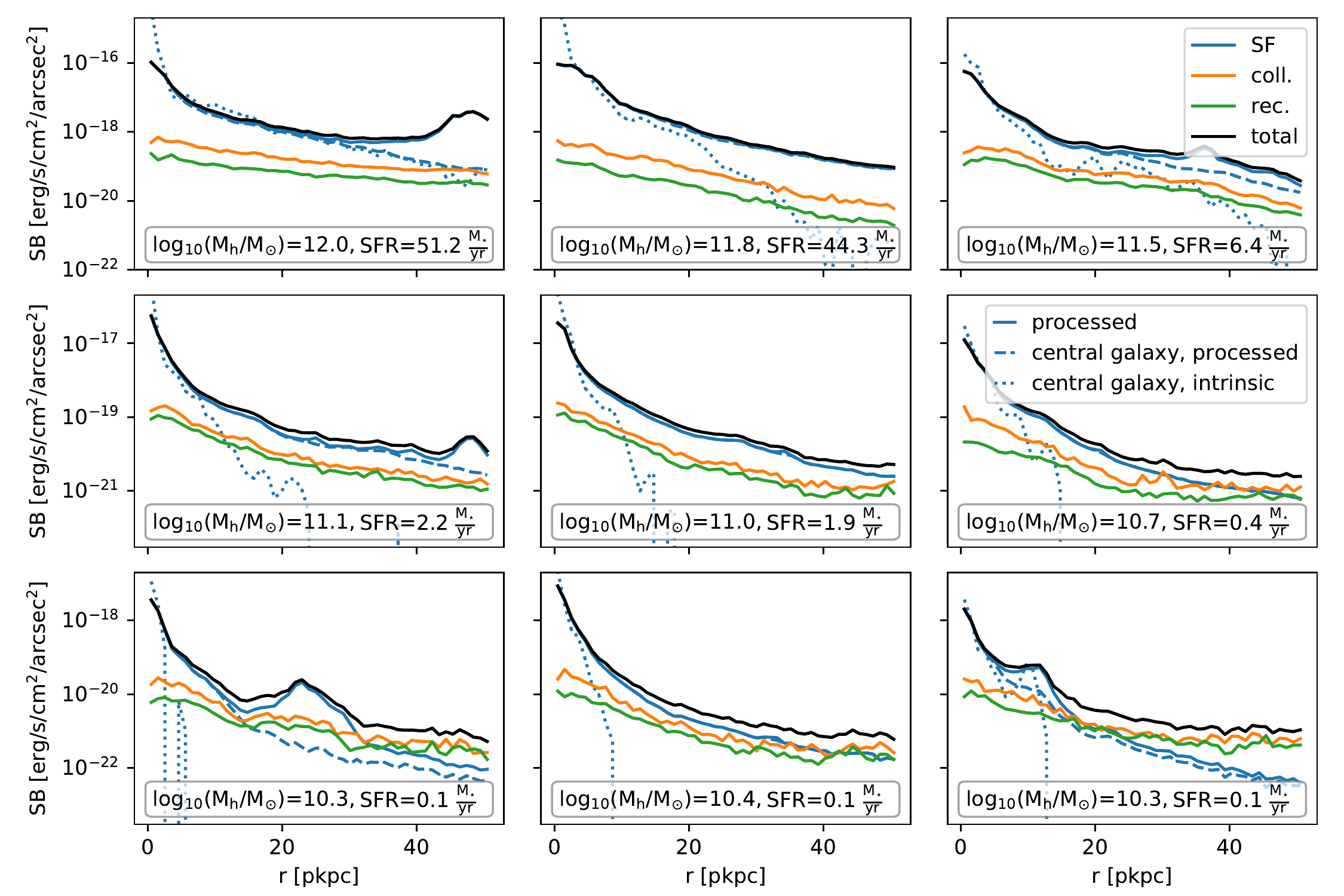}
\caption{Radial surface brightness profiles of nine individual Lyman-$\alpha$ halos at $z=3$ in TNG50. The objects and plotting order are the same as in Figure \ref{fig:panel3x3_total_z3_2Dviews}. The black solid lines show the total, observable radial profiles extracted from the surface brightness maps of processed photons. We decompose these profiles into their three physical emission sources, namely star-formation (SF; blue), collisional excitations (coll; orange) and recombinations (rec; green). Generically, the radial profiles drop steeply within $\lesssim 5\ $pkpc and flatten towards larger radii. For star-formation, we also split profiles into intrinsic (processed) photons from the central galaxy only as dotted (dashed) blue lines, emphasizing the important and non-negligible role of scattering in outer LAHs.}
\label{fig:panel3x3_contributions_z3_radialprofiles}
\end{figure*}

In Figure \ref{fig:panel3x3_total_z3_2Dviews} we show surface brightness maps for a collection of nine LAHs ordered by star-formation rate at $z=3$, including the two LAHs from Figure \ref{fig:overviewplot}. The three colored contours all trace a surface brightness of $10^{-19}\ $erg/s/cm$^{2}$/arcsec$^{2}$, differentiating between the observable flux (i.e. from processed photons; blue), intrinsic emission (red), and intrinsic emission due to star-formation alone (yellow).

Broadly, the Ly$\alpha$ scattering process increases the apparent sizes of LAHs beyond that of intrinsic emission. In some instances, such as the lower right panel, red contoured regions occur without corresponding yellow contours, indicating emission without the presence of star-formation in filamentary structures. We note that the surface brightness distributions of the scattered Ly$\alpha$ photons are significantly smoother and isotropic than the more complex structure of the underlying density field. Qualitatively this is compatible with the low eccentricities found for LAHs in observations~\citep{Wisotzki2016}. At distances beyond $\gtrsim 20\ $pkpc from the central galaxies, LAHs become increasingly anisotropic, due to the combination of satellite galaxies (for the more massive halos) and anisotropic gas inflows.

\subsection{Radial Profiles}

In Figure \ref{fig:panel3x3_contributions_z3_radialprofiles} we show the radial Ly$\alpha$ surface brightness profiles extracted from the same nine intensity maps of Figure~\ref{fig:panel3x3_total_z3_2Dviews}. We decompose the total profile (including rescattering; black) into its contributions from the three emission sources: star-formation (solid blue), collisional de-excitation (orange), and recombination (green). For star-formation, we additionally show the emission neglecting subhalo/satellite contributions, both intrinsic (dotted blue) and processed (i.e. scattered, dashed blue).

In general, radial profiles are steep near the center of the halo and quickly flatten beyond $r\gtrsim 15\ $pkpc. The inner regions are dominated by star-formation sourced recombination, and even at larger radii up to $50\ $pkpc, star-formation often remains the dominant contribution while excitations and recombinations (orange and green lines) reach similar magnitude.
Except for the most massive halos in the upper row, intrinsic \Lya emission from star-forming regions (blue dotted line) quickly fades within $r\lesssim 15\ $pkpc. At larger radii, the profiles are shaped by scattered photons. However, as expected, there are occasional bumps in the profile from star-formation in satellite galaxies of more massive halos.

At larger radii, profiles are shaped by scattered photons, highlighting two important effects of the radiative transfer. First, the central surface brightness is severely damped by photon rescattering (dashed blue vs. dotted blue curves). Second, large amounts of those Ly$\alpha$ photons are scattered further out, which provides an important contribution of star-formation to the extended radial profiles even though little in-situ star-formation may take place at those radii.

Collisional excitations and recombinations become important at radii $r\gtrsim 20\ $pkpc for low mass halos, and the typical surface brightness contribution from collisional excitation exceeds that from recombination by a factor of $\sim 2$. The most massive galaxies more frequently host nearby satellites, which result in the occasional bumps in the profile from star-formation in these subhalos.

\begin{figure*}
\centering
\includegraphics[width=0.94\linewidth]{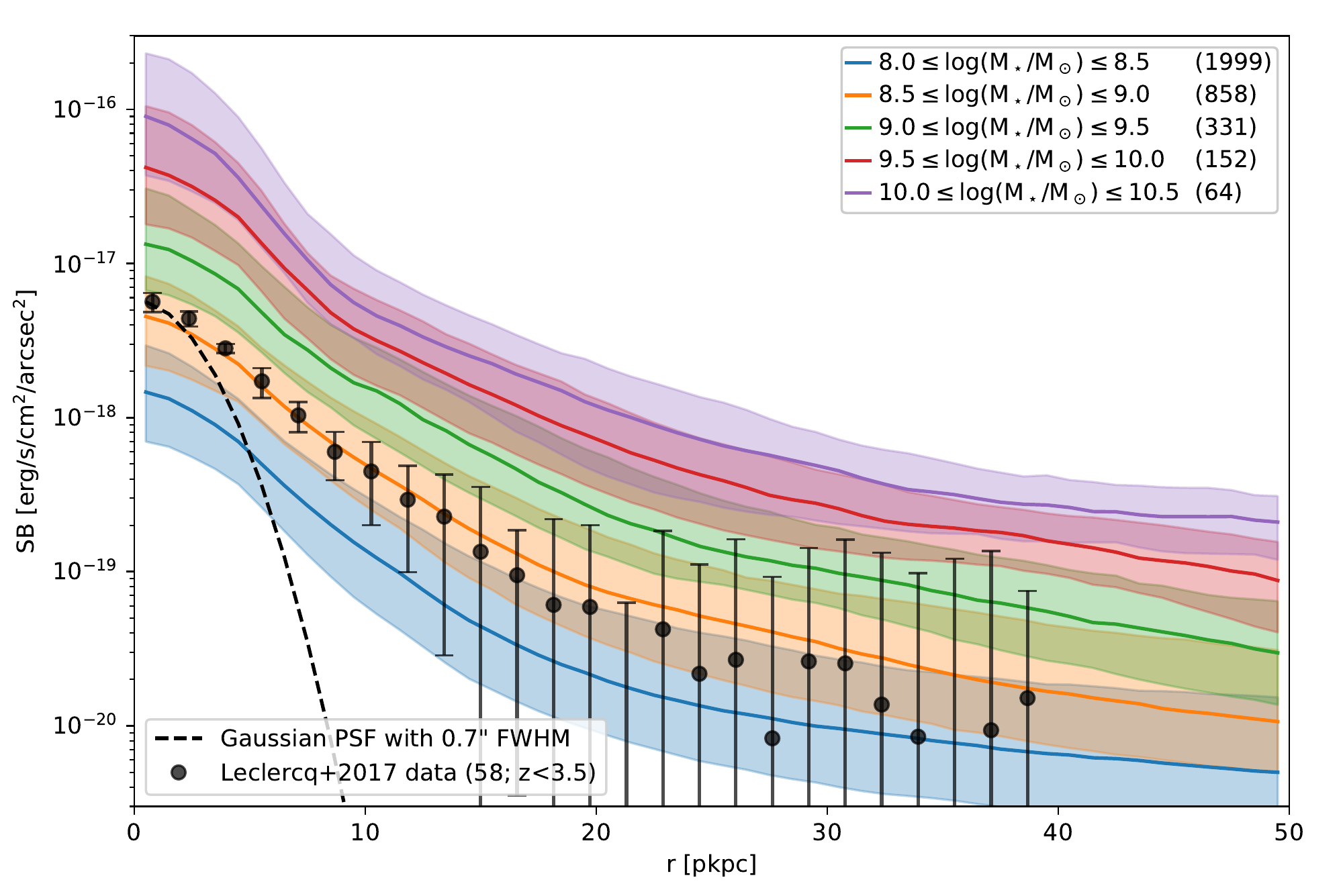}
\includegraphics[width=0.48\linewidth]{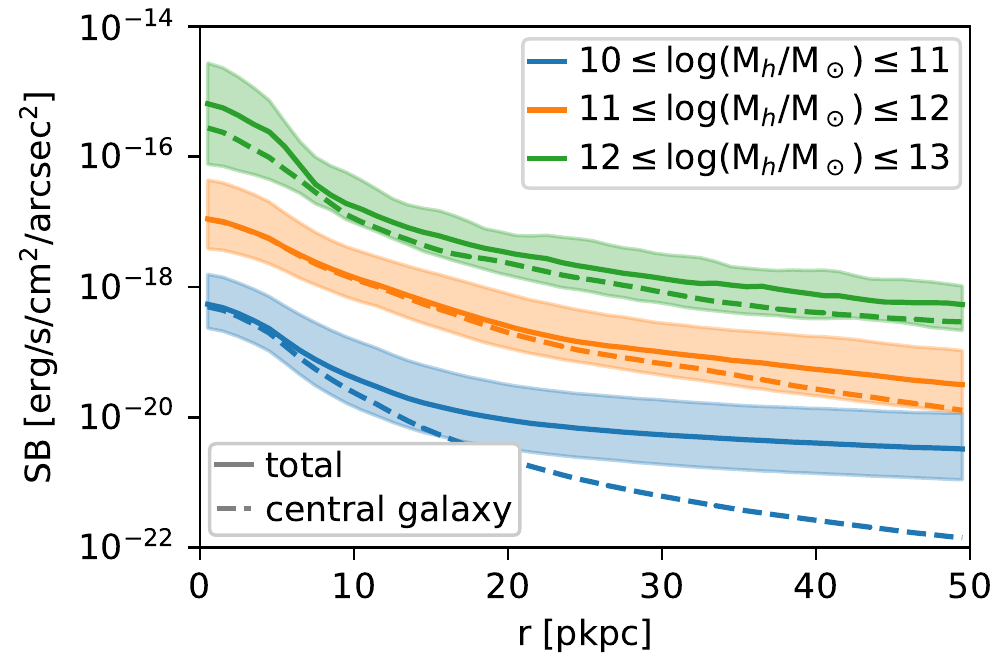}
\includegraphics[width=0.48\linewidth]{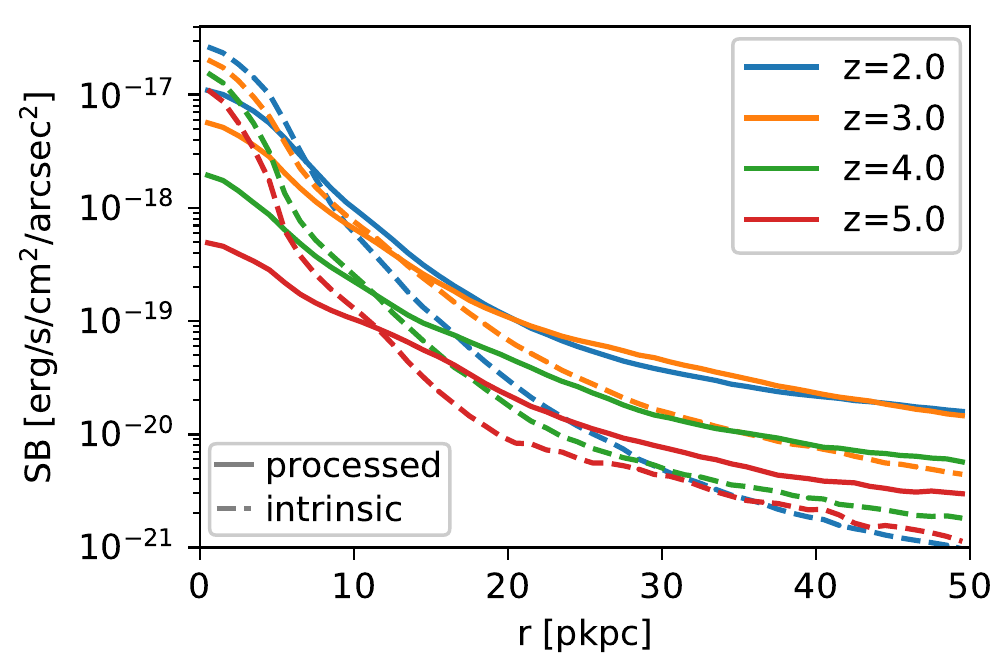}
\caption{Overview of the predicted median Ly$\alpha$ surface brightness profiles around TNG50 galaxies, as a function of mass and redshift. \textbf{Main panel}: Stacked radial profiles for five different stellar mass ranges from $M_\star = 10^8$\msun to $10^{10.5}$\msun, at $z=3$. The respective count of contributing halos to the stack is given in parentheses in the legend. At fixed cosmic time, the central Ly$\alpha$ surface brightness increases monotonically with stellar mass, roughly as $\rm{SB}(\rm{r=0}) \propto M_\star^{0.9}$. We overplot the median radial profile of 58 observed LAHs at redshifts between $2.9$ and $3.5$ presented in \protect\cite{leclercq17} from the MUSE UDF (see text for details). \textbf{Lower left}: Stacked radial profiles as a function of dark matter halo mass, also at $z = 3$. The solid lines show the observed radial profiles, while the dashed lines only consider those photon contributions originating from the central halo. \textbf{Lower right}: Redshift evolution of stacked radial profiles for a fixed galaxy stellar mass bin of \smassdflt. At fixed stellar mass, Ly$\alpha$ halos are more luminous towards lower redshift. We contrast the full radiative transfer result (RT; solid lines) with the intrinsic emission profiles (dashed lines). The scattering which occurs during the RT lowers the Ly$\alpha$ surface brightness at halo center ($\lesssim 5$ pkpc) while increasing it at larger radii, producing an overall flatter profile.}
\label{fig:rprofiles_stacked}
\end{figure*}

We move from the case study of individual profiles to a quantitative exploration of the average predicted LAH profiles. In Figure \ref{fig:rprofiles_stacked} we stack galaxies based on stellar mass (main panel) and halo mass (lower left panel) at $z=3$. We also show the evolution with redshift from $z=2$ to $z=5$ in a fixed stellar mass bin (lower right panel). Profiles always show the median stacked profile after radial binning, which we note is not the same as first median stacking the two-dimensional surface brightness maps. For the radial binning we calculate the mean surface brightness for a given annulus.
Shaded regions show the central 68 percentiles. The dashed line in the main panel indicates the $0.7$\,arcsec FWHM Gaussian PSF we adopt, which dominates the smoothing of the radial profiles at  small distances $r < 10\ $pkpc. At larger radii, the surface brightness rapidly flattens, as we explore below.

In the top panel, stacking surface brightness in stellar mass bins from low-mass galaxies with $M_\star = 10^8$\msun to Milky Way mass systems with $M_\star = 10^{10.5}$\msun, we see that Ly$\alpha$ surface brightness increases monotonically, at all radii, with increasing stellar mass. Despite this strong correlation between peak surface brightness and stellar mass, the overall shape of the median radial profiles is largely independent of the stellar mass. We note that a mass-dependent \Lya photon escape probability, e.g. due to dust physics, would impact the trend of overall luminosity and stellar mass, which we explore further in Section \ref{sec:centralbrightness}.

The lower left panel of Figure \ref{fig:rprofiles_stacked} shows the radial profiles in three halo mass bins. The central surface brightness of the Ly$\alpha$ profiles rises as a function of halo mass. In addition, we more clearly find a change in shape of the radial profiles as a function of halo mass, whereby flattening begins at smaller radii for lower mass halos. Considering central galaxy emission only (dashed lines), we observe that as star-formation decreases towards lower mass, the luminosity budget available for rescattering in the halo does likewise. Hence, diffuse emission outside of the central galaxy becomes dominant at smaller radii. Equivalently, since the central (or total) luminosities are lower at lower mass, external emission (i.e. from other halos) can take over more quickly.

In the lower right panel of Figure \ref{fig:rprofiles_stacked} we explore the redshift evolution of radial profiles in a fixed stellar mass bin of \smassdflt. Towards higher redshift, the central surface brightness drops significantly while changes to the overall shape are minor. Here we also show the radial profiles of the intrinsic photons (dashed lines), where Ly$\alpha$ photons are allowed to escape directly to the observer without scattering, contrasting against the processed emission (solid lines). The intrinsic profiles' lower brightnesses at higher redshifts are driven by surface brightness dimming, which is however largely countered by the higher specific star formation rates at fixed stellar mass. The intrinsic central SB luminosities decrease much more slowly towards higher redshift, implying that the photon redistribution due to resonant scattering is significantly more important at higher redshifts.

\subsection{Comparison of TNG50 and MUSE data}
\label{sec:MUSEcomparison}

In the top panel of Figure \ref{fig:rprofiles_stacked} we also overplot the median radial profile from the observational LAH dataset of \cite{leclercq17} for objects with redshifts between $2.9$ and $3.5$. The data set is based on the MUSE Ultra Deep Field (UDF), which finds extended Ly$\alpha$ emission around 145 of 184 star-forming galaxies at $3 \lesssim z \lesssim 6$ with a median of $z\sim 3.7$. The observed galaxies extend in stellar mass down to $\sim 10^7$\msun with an average stellar mass of $M_\star \sim 10^{8.5}$\msun \citep{boogaard18}. As a result, the most appropriate comparison is against the orange line, where we find a excellent agreement of the normalization and radial shape of the surface brightness profile between MUSE and TNG50. At larger radii $r > 20$\,pkpc the observed stacked profile becomes uncertain given the large errorbars, so we cannot ascertain whether or not the strong flattening we observe in the TNG50 LAHs is also seen in the data. Additionally, this flattening is affected by MUSE's more extended Moffat PSF~\citep{Bacon2017}. The agreement of stacked radial profiles between observed and simulated samples degrades toward higher redshifts, particularly $z=5.0$. This discrepancy is largely driven by the flattening of the simulated profiles at high redshifts (see bottom right panel) that is not found in the observed sample.

We note that quantitatively comparing TNG50 to Lyman-$\alpha$ halo observations has many challenges and subtleties. With respect to the MUSE data of \cite{leclercq17} in particular, although we address the issue of spatial smoothing and the PSF, remaining systematics could include details of (i) spectral smoothing, resolution,  and the construction of the surface brightness map with a frequency-space integration; (ii) noise, including relevant surface brightness limits and sky backgrounds; (iii) the position chosen as the center of the halo; (iv) the derivation of galaxy properties, including stellar mass, where SED fitting based on HST broad-band photometry lacks near-IR points, and neglecting emission lines including Ly$\alpha$ which can fall into the F606W filter \citep{feltre20}; and (v) sample selection and selection biases, i.e. choosing appropriate analogs for comparison to the observed halos and/or matching the observed galaxy population in general.

\begin{figure*}
\centering
\includegraphics[width=0.99\linewidth]{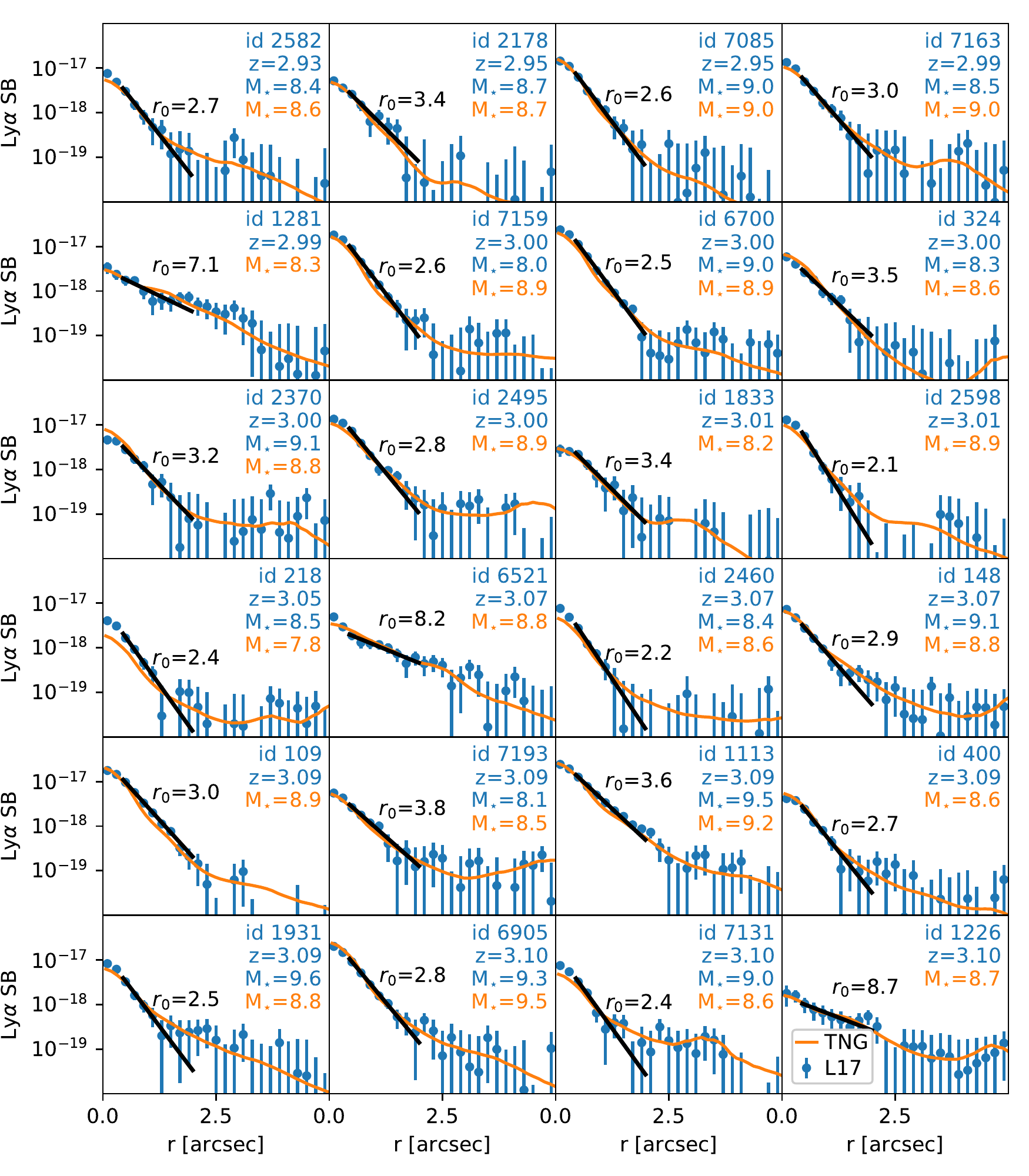}
\caption{Gallery of observed Lyman-$\alpha$ halos from the MUSE UDF \protect\citep[blue points with uncertainties;][]{leclercq17}, chosen as the 24 closest to redshift $z=3$. Every observed LAH is matched to a simulated halo from TNG50 by choosing the best least-squares fit profile. All simulated radial profiles are smoothed with a Gaussian $0.7\ $arcsec FWHM PSF, and overplotted (orange lines). This demonstrates that the simulation has the diversity and sample statistics to recover excellent matches to all observed halos, and that TNG50 can successfully reproduce every observed profile, in both normalization and shape, with at least some simulated halo. For each halo, we include the ID and redshift of the observed MUSE object, as well as its stellar mass estimate if available \protect\citep{feltre20}. We also show the exponential scale length $r_0$ (in pkpc) fitted to the MUSE data. For the simulated profile we also include its galaxy stellar mass: observed LAHs and their simulated matches from TNG50 surround galaxies of similar mass.}
\label{fig:lec0}
\end{figure*}

Beyond the stacked profile comparison, we also contrast individual LAHs as observed in the MUSE UDF to those from TNG50. Figure \ref{fig:lec0} shows the 24 radial profiles from \citet{leclercq17} closest in redshift to $z=3$ (blue data points with errorbars). For each, we search for the best `match' from among our catalog of simulated LAHs at that redshift, and select the single LAH with the minimum least-squared difference\footnote{Note that the low surface brightness measurements at large radii \mbox{$r > 2$\,"}, have only small weights in the linear least-square fit.} which are overplotted (orange).

\begin{figure*}
\centering
\includegraphics[width=0.99\linewidth]{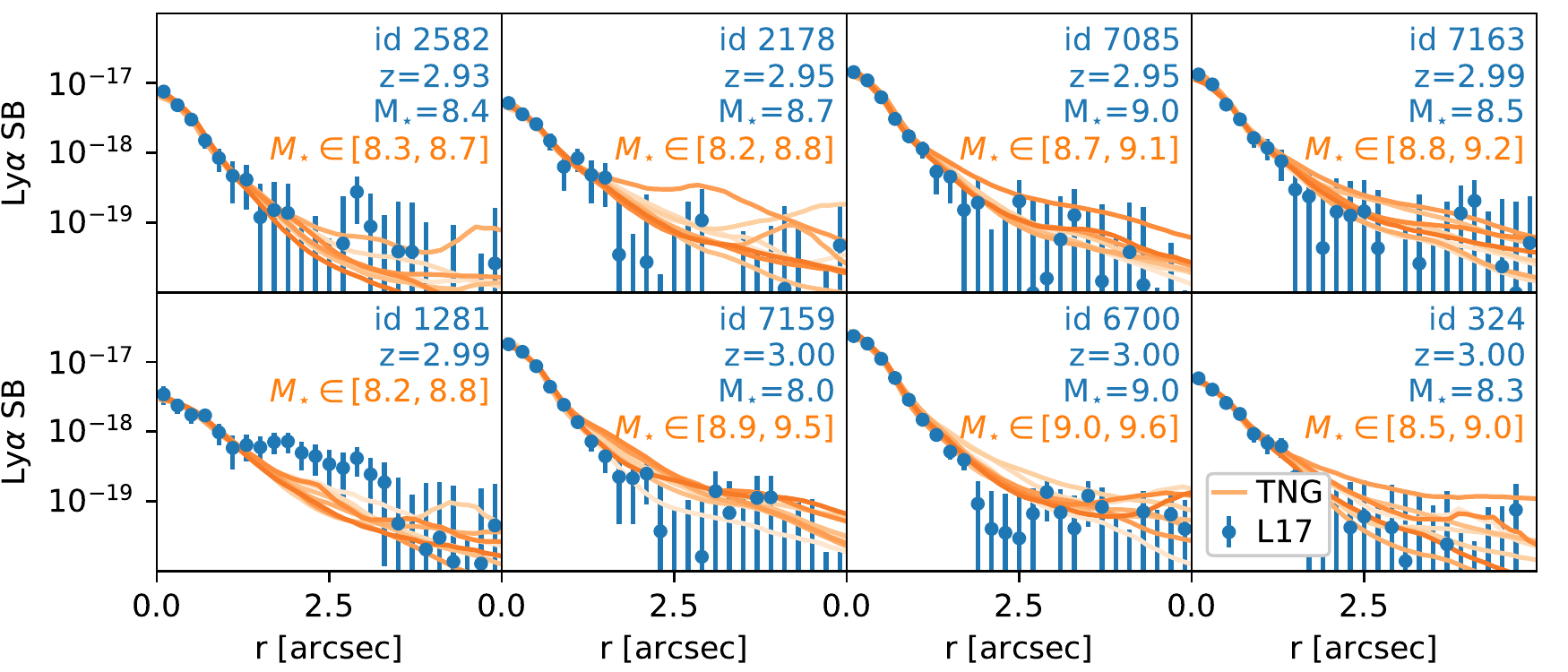}
\caption{Gallery of observed Lyman-$\alpha$ halos from the MUSE UDF \protect\citep[blue points with uncertainties;][]{leclercq17}, where here we show the first 8 LAHs shown from Figure \ref{fig:lec0}. As before, we search for the best matching LAH from TNG50, but now only fit data points for small radii $r<2\ $arcsec, and show the best five fits for each profile. As before, we contrast the MUSE stellar mass (if available) with the stellar mass range of the five simulated profiles, which are shaded from light to dark orange with increasing stellar mass. We find that although the simulated profiles show object-to-object variances at large distance, in most cases they are consistent with the observed profile within its errorbars.}
\label{fig:lec0_2arcsec}
\end{figure*}

In general, we are able to find excellent matches to the observed data, demonstrating that TNG50 can reproduce the diversity and variety of observed LAH profile shapes. For instance, we show good matches for more compact and more extended objects (e.g. id 2178, id 6521), and although noise starts dominate at larger radii, we also find good matches for candidates with very flattened profiles (e.g. id 1226). The stellar masses of the observed galaxies, and the stellar mass of the selected TNG50 analog, are both shown in the legend for reference. For the subsample of observed LAHs for which we have a stellar mass estimate \citep[from][]{feltre20}, we find a mean difference in $M_{\star,\rm{MUSE}} / M_{\star,\rm{TNG}}$ of $0.11$ dex with a standard deviation of $0.43$ dex. This indicates that the simulated LAHs selected as good matches surround galaxies with comparable stellar masses as the observed systems.

Figure \ref{fig:lec0_2arcsec} shows a similar comparison of matched LAH profiles between the MUSE data and TNG50. However, we now restrict the least-square fit to those values at distances below $r\leq 2\ $arcsec, and include the five matches in each case to highlight the range of predicted large radii behaviour from simulated profiles. Most of those fits are compatible with the observed radial profiles and their error-bars at large radii \citep[see also][for fits to L17 SB(r) profiles using 3D Ly$\alpha$ RT coupled to idealized shell models rather than cosmological simulations]{song20}. A notable exception is the profile of MUSE-id 1281 which has an excess at $r \sim 2\ $arcsec, possibly due to the existence of a satellite galaxy.

\subsection{The Origin and Source of Lyman-alpha Halo Photons}

\begin{figure}
\centering
\includegraphics[width=1.0\linewidth]{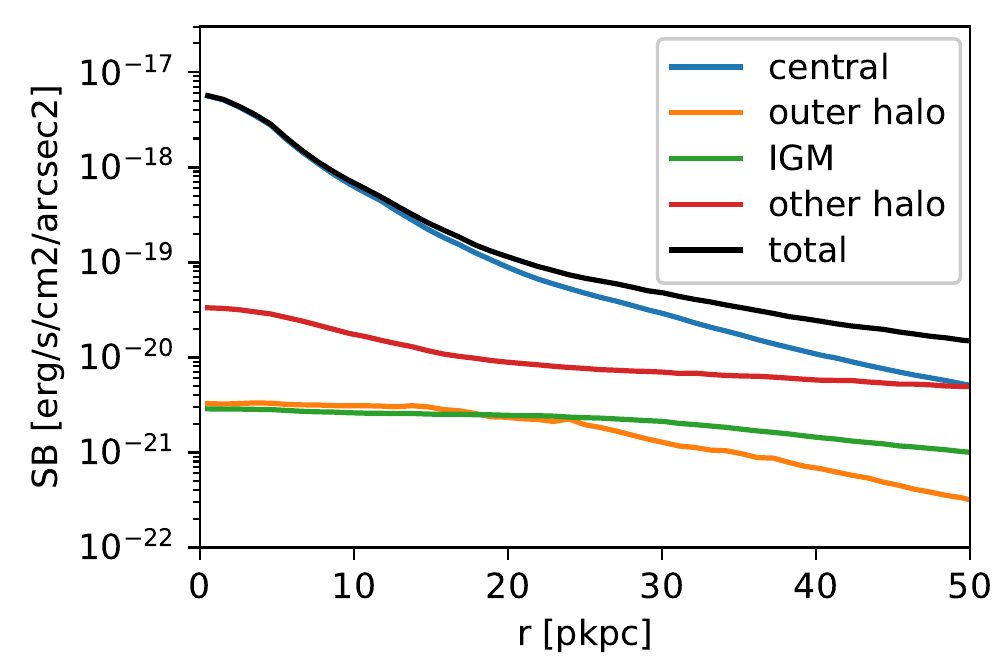}
\caption{Median stacked radial Ly$\alpha$ profile for galaxies with stellar masses \smassdflt{} at $z=3$ in TNG50. We decompose this profile into the photons with differing emission origins: from the central galaxy (blue), outer halo (orange), the IGM (green), and other halos (red). At $r<20\ $pkpc emission from the central subhalo dominates, beyond which contributions from other halos start contributing to the overall shape. Beyond $40\ $pkpc the `other halo' origin is critical and produces the flattening of the profiles towards large radii. The contributions from emission originating in the outer parts of the halo and the IGM are negligible.}
\label{fig:rprofiles_origins}
\end{figure}

Although LAHs are observed localized around galaxies and their dark matter halos, the photons which contribute to that emission can arise from a number of different origins. In Figure \ref{fig:rprofiles_origins} we show the relative contribution to a stacked Ly$\alpha$ surface brightness profile, depending on the origin of emission, for galaxies in the stellar mass bin \smassdflt. We categorize the emission origin of each photon as one of the four categories previously introduced in Section \ref{sec:postprocessing}.

We find that emission from the central galaxy (blue) clearly dominates the radial profile below $r\lesssim 20\ $pkpc after which emission originating from other halos but scattering onto the targeted halo becomes increasingly important. Above $40\ $pkpc this `other halo' origin (red) even dominates the radial profiles. The radial profile from rescattered photons originating in other halos has a very shallow slope, thus leading in large part to the flattening of the overall profiles at larger radii. In particular, we find that the contribution from other halos is significantly boosted if more massive halos are nearby, an effect we explore more in Section \ref{sec:largescales}.

Emission originating in the IGM (green) and outside of the central subhalo (particularly from satellites; orange) is generally negligible, and never contributes more than a few percent to the total stacked profile. However, there can be infrequent radial profiles of individual halos with larger contributions from IGM and satellites than for the shown average. We note that the IGM contribution in particular will depend upon the line-of-sight integration depth.

For higher mass halos (not shown) we find that the emission from the central subhalo grows more rapidly than any of the other origins, pushing the observed flattening to larger radii. Analogously, lower mass halos flatten at smaller radii. Other halos start to significantly contribute ($\geq$ 10\% to the total stacked profile) at $7$, $22$ and $\geq 50\ $pkpc\footnote{In the highest mass bin, the radius lies outside of the $50\ $pkpc radius aperture.} for the respective 1 dex stellar mass binned halos starting at $7.5$, $8.5$ and $9.5$ $\log\left(M_\star / \rm{M}_\odot\right)$. There is very little redshift evolution for these radii from $z=2$ to $z=5$.

\begin{figure}
\centering
\includegraphics[width=1.0\linewidth]{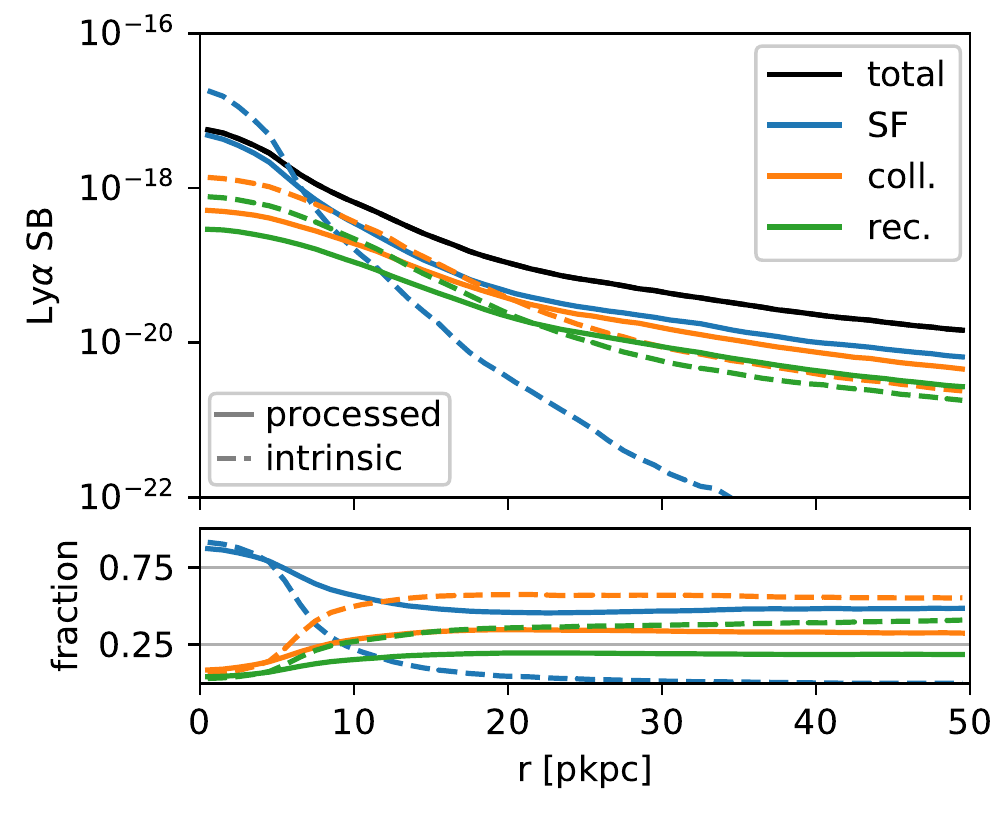}
\caption{Median stacked Ly$\alpha$ surface brightness radial profiles at $z=3$ for halos with \smassdflt{} decomposed into different emission sources (upper panel), and the relative fraction of each (lower panel) in TNG50. Dashed lines show intrinsic emission, while solid lines show the processed (i.e. scattered) signal. Emission from star-forming regions typically dominates the intrinsic emission up to $10\ $pkpc after which collisional excitations start to dominate. However, radiative transfer redistributes this central emission towards the halo outskirts, such that star-formation remains the dominant source of Ly$\alpha$ emission for observed LAHs at all radii shown.}
\label{fig:rprofiles_mechanisms}
\end{figure}

In Figure \ref{fig:rprofiles_mechanisms} we similarly decompose the stacked profile into the relative contributions of different emission \textit{sources}: star-formation sourced rescattered photons (blue), collisional excitation (orange), and recombination (green). In addition to the `processed' signal (solid lines) we also show the intrinsic emission signal (i.e. ignoring scattering effects; dashed lines). In both cases we find that star-formation makes up the bulk of the SB within the central $10\ $pkpc. At larger radii however rescattered photons from star-forming regions drop to a $\sim 50\%$ relative contribution as diffuse collisional excitations and recombinations rise to $\sim 30\%$ and $\sim 20\%$ respectively. 
These relative fractions, shown here at $z=3$, are similar at other redshifts (not shown).

\subsection{Lyman-alpha Halo Sizes}
\label{sec:sizes}

\begin{figure}
\centering
\includegraphics[width=1.0\linewidth]{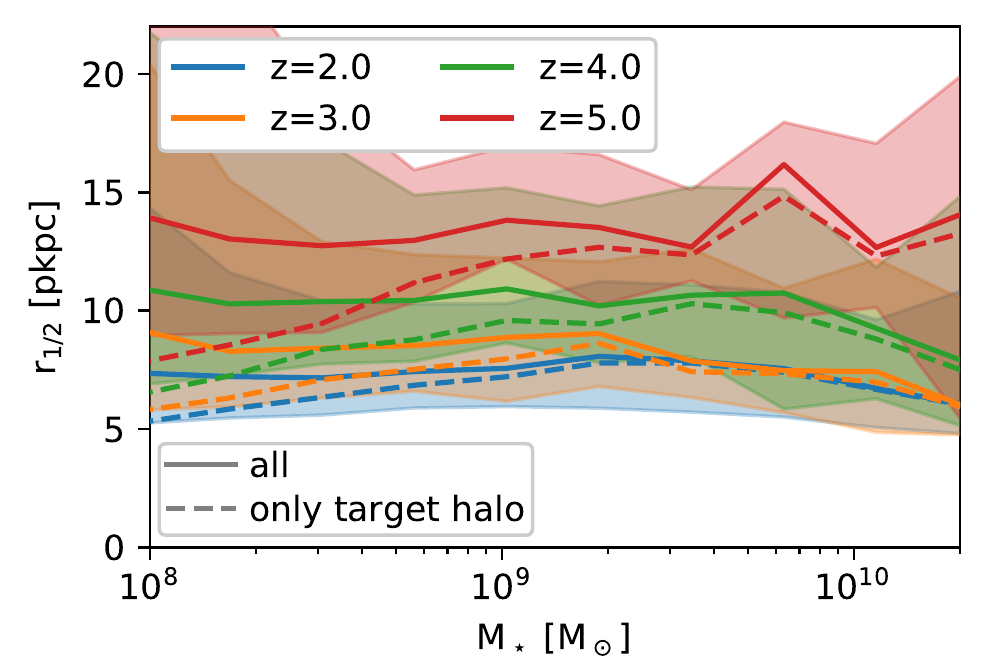}
\caption{The \Lya half-light radius $\rm{r}_{1/2}$ as a function of stellar mass over redshift in TNG50. The colored solid lines show the median radius and the shaded regions enclose the radii within the central $68\ $ percentiles. The typical $r_{1/2}$ size lies between $5$ and $15\ $pkpc, and monotonically increases with redshift, while being mostly constant over the stellar mass range. For the colored dashed lines, contributions originating outside of the targeted halo have been ignored. The difference between solid and dashed lines therefore indicates an increasing impact of unbound and other halos' contributions at low stellar masses.}
\label{fig:rprofiles_rhalf}
\end{figure}

To study the dependence of LAH profiles on galaxy/halo properties and redshift, we summarize each profile by a characteristic size and surface brightness, as defined in Section \ref{sec:reducedstats}. Figure \ref{fig:rprofiles_rhalf} shows the Ly$\alpha$ half-light radius $r_{1/2}$ as a function of stellar mass across the studied redshift range. We find that the half-light radii of our LAHs is typically between $5$ and $15\ $pkpc. There is little correlation with stellar mass compared with the scatter. Half-light radii are systematically larger in physical kpc towards higher redshift, roughly doubling in size from $z=2$ to $z=5$ across the redshift range. This implies a stronger redistribution through scatterings into the outskirts of halos at higher redshifts, which is largely driven by a higher neutral hydrogen density at those redshifts.

\begin{figure}
\centering
\includegraphics[width=0.99\linewidth]{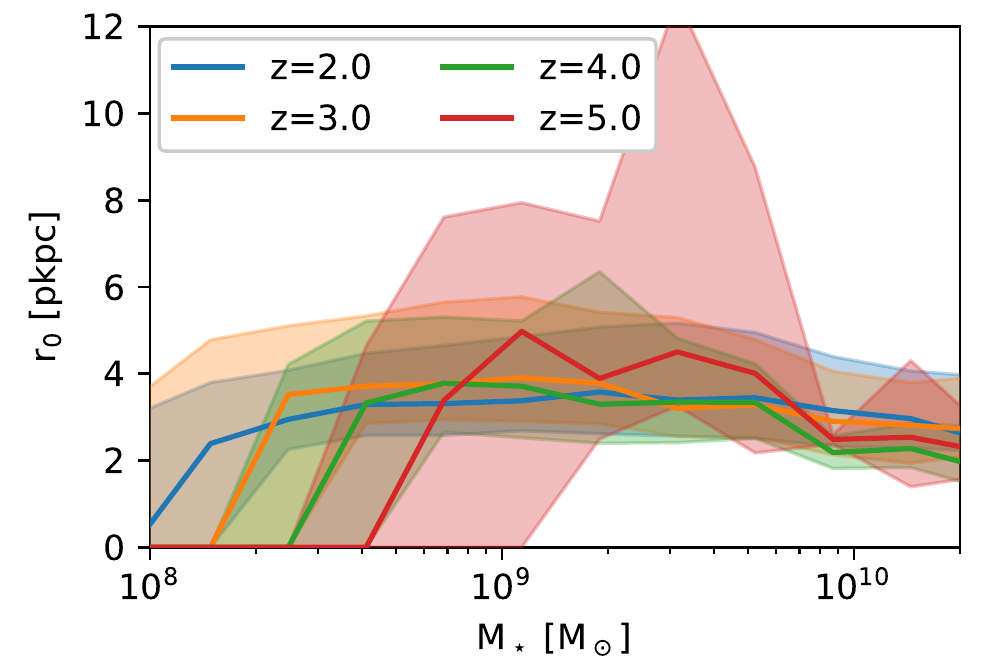}
\includegraphics[width=0.99\linewidth]{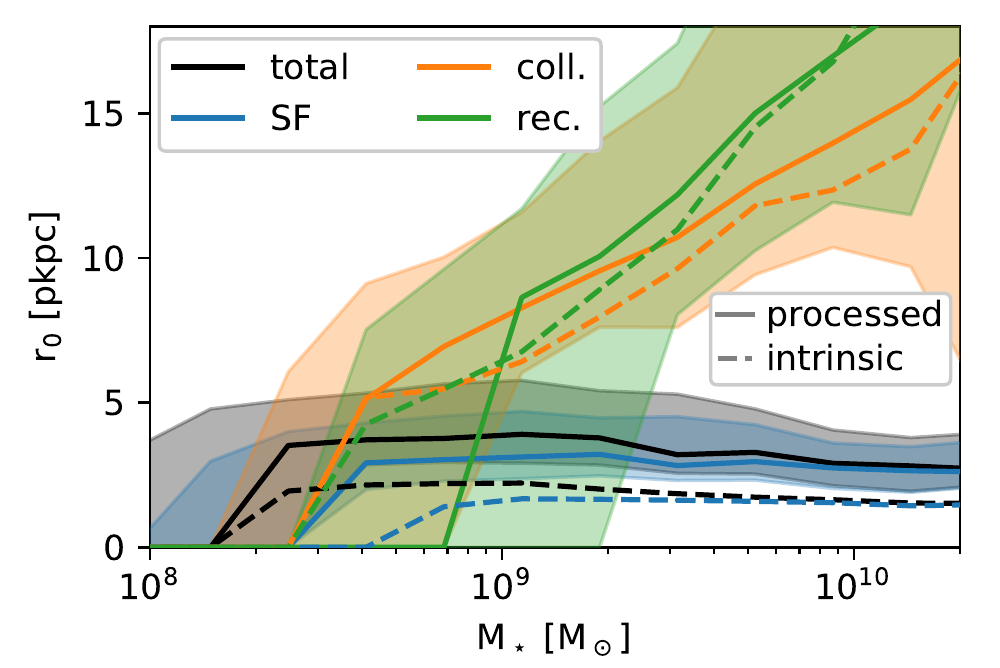}
\includegraphics[width=0.99\linewidth]{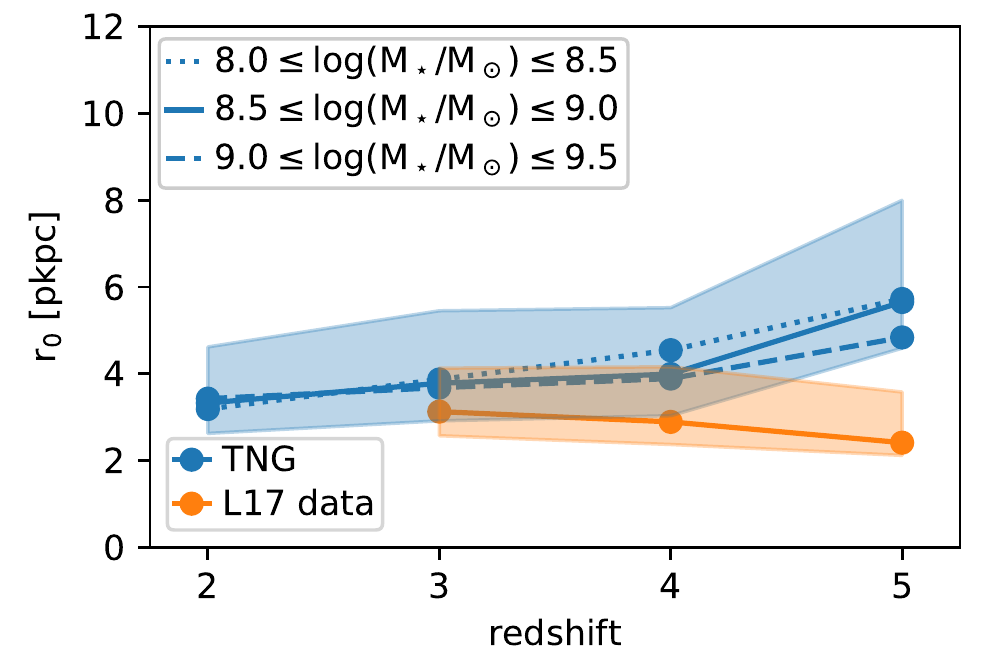}
\caption{The \Lya single exponential radius $\rm{r}_{0}$ as a function of stellar mass and redshift. The colored solid lines show medians, while shaded regions indicate the central $68$ percentiles. A drop of the scale radius to zero at around $\log_{10}\left(M_\star/M_\odot\right)\sim 8.0-8.4$ is given due to the noise modeling (see text).
\textbf{Top}: $r_0$ as a function of stellar mass, at four distinct redshifts. There is no clear trend of $r_0$ with either redshift or stellar mass. 
\textbf{Middle}: Decomposition of $r_0$ at $z=3.0$ (black) into its three emission origins (SF, coll, rec), for both intrinsic and processed photons. As the luminosity budget in the proximity of the halos' center is dominated by star-formation, the latter effectively sets the scale length. 
\textbf{Bottom}: $r_0$ versus redshift compared to observational data from the MUSE UDF \protect\citep{leclercq17}. For simulations and observations, we fit the scale length $r_0$ using the same procedure. We show three different bins of fixed stellar mass for TNG50 (blue lines). At fixed stellar mass, LAH sizes are overall larger towards higher redshift. No clear redshift trend is evident in the observations which are consistent with no size evolution, although the galaxy stellar mass distribution as a function of redshift in the data is uncontrolled (see text for details).}
\label{fig:rprofiles_expR0}
\end{figure}

In Figure \ref{fig:rprofiles_expR0} we show the fitted single exponential scale length $r_0$. In all panels we impose the noise modeling described in Section \ref{sec:reducedstats}, which effectively imposes a sensitivity limit and induces a lower cut-off mass below which modelled LAHs show no \textit{observable} extent. This limit starts to affect the median at a stellar mass below $\sim 2 \times 10^8\ $M$_\odot$ ($\sim 10^9\ $M$_\odot$) at $z=2$ ($z=5$), causing the artificial drop-offs at low $M_\star$.

Except for this sensitivity limit, there are no clear trends of $r_0$ with stellar mass. Similarly, no clear redshift evolution is evident (upper panel). However, when looking at our fiducial stellar mass range of \smassdflt{} only, we do find that $r_0$ increases with redshift from $3.4_{-0.7}^{+1.3}\ $pkpc at $z=2$ to  $3.7_{-0.9}^{+1.7}\ $pkpc ($4.0_{-1.0}^{+1.7}\ $pkpc, $5.5_{-1.6}^{+2.4}\ $pkpc) at $z=3$ ($z=4$, $z=5$).

The middle panel of Figure \ref{fig:rprofiles_expR0} shows the scale radius as a function of stellar mass at $z=3$. Here we decompose the contribution to LAH size by emission source, by determining $r_0$ based on each of three respective emission sources alone (colored lines). The black solid line shows the median $r_0$ relation combining all three emission sources, as would be observable. We find that the overall scale radius $r_0$ is largely determined by the emission from star-formation due to its high surface brightness. Scale lengths from star-formation show no correlation with stellar mass, while both collisions and recombinations do show a strong positive correlation. Thus, $r_0$ follows the lack of evolution with mass seen in the star-formation source. In contrast, both diffuse collisions and recombinations show a strong positive correlation with mass. For halos between \smassdflt{} we find $r_0$ to be $3.7_{-0.9}^{+1.7}\ $pkpc for the overall profile and $3.1_{-0.7}^{+1.4}\ $pkpc, $8.0_{-1.7}^{+2.8}\ $pkpc or $9.2_{-2.0}^{+3.2}\ $pkpc for SFR, excitations, recombinations only respectively. 

Solid lines indicate processed photons, while the dashed lines show $r_0$ based on the intrinsic photons only. Not surprisingly, intrinsic photons typically give rise to a smaller scale length than processed photons. Most importantly, intrinsic photons from star-formation give rise to a scale length close to a point-like source convolved with the PSF, while this value doubles from rescattering in the CGM.

In the bottom panel of Figure \ref{fig:rprofiles_expR0} we show the explicit redshift evolution of LAH sizes, from $z=2$ to $z=5$, and compare the TNG50 result with that of \citet{leclercq17}. For the comparison with MUSE, we show simulated halos in stellar mass bins $\log_{10}\left(M_\star/\rm{M}_\odot\right)$ from $8.0$ to $9.5$, which overlap the observed distribution. At fixed stellar mass, we find that the median $r_0$ increases towards higher redshift and the magnitude of this trend is similar to the scatter of $r_0$ at each redshift. For the MUSE UDF observations, we find $3.1_{-0.6}^{+1.0}\ $pkpc, $2.9_{-0.5}^{+1.3}\ $pkpc and $2.4_{-0.3}^{+1.2}\ $pkpc for the redshift bins centered at $z=3$, $z=4$ and $z=5$, respectively. 

Our face value comparison with the MUSE data implies that LAHs in TNG50 are slightly more spatially extended than in observations, by roughly a factor of $\sim 1.2$ at redshift three, increasing to $\sim 2.3$ by redshift five with a simulated $r_0\sim 5.5\ $pkpc, the two size distributions being roughly one standard deviation apart. We caution, however, that the stellar mass distribution of the observed galaxies is not fixed as a function of redshift. More massive galaxies are more easily observed at higher redshift, and ideally we would match the joint $(M_\star,z)$ distribution to make this comparison. However, remaining methodological differences likely still dominate the uncertainty in this comparison, as sizes are not measured in exactly the same way in both the observational data and TNG50 simulation.

We note that observed sources of Ly$\alpha$ emission have been characterized by many different sizes ranging from roughly $2\ $pkpc to $9\ $pkpc at $z\sim 3$ \citep{Bond09,Momose2014}. This large range of radii hints at the large diversity of different galaxy selection (LBGs/LAEs), functional fitting forms (single/double exponential) and methodologies (individual/stacking). We leave a quantitative comparison of \Lya sizes for future work, and do not try to explicitly compare LAE sizes to the observational literature here.

\subsection{LAH Central Brightness}
\label{sec:centralbrightness}

\begin{figure}
\centering
\includegraphics[width=1.0\linewidth]{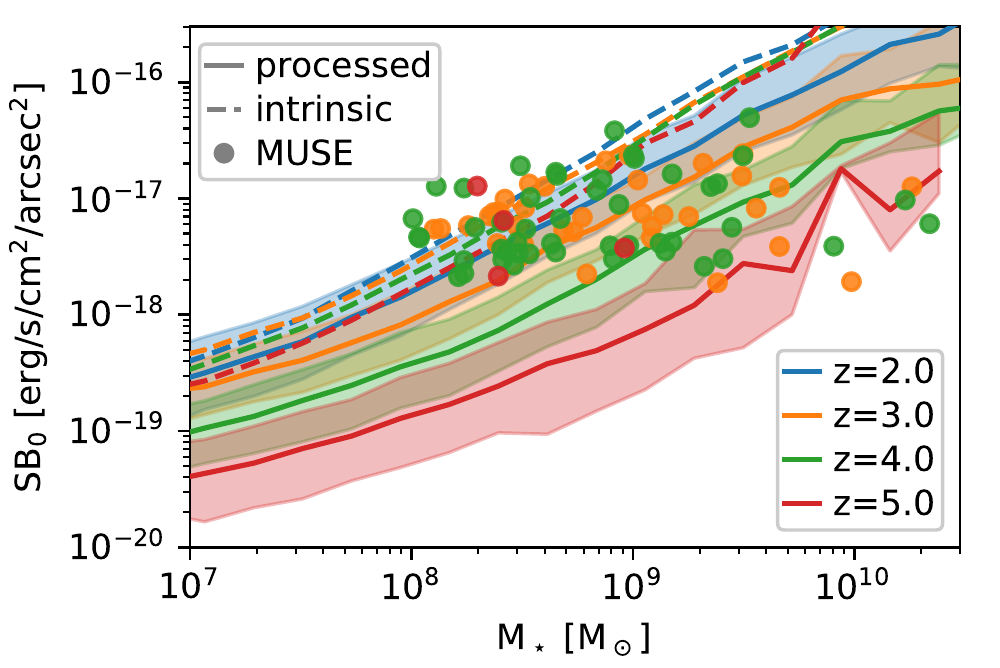}
\caption{The central surface brightness SB$_0$ as a function of stellar mass across the studied redshift range in TNG50. The solid lines show the median for the processed photons for a given mass bin and the shaded region shows the corresponding 68 central percentiles. In dashed, we show the median for the intrinsic photons. The dots show measured MUSE results color coded by simulated redshift they are closest to. The central surface brightness is a strong function of galaxy mass both for intrinsic and processed photons in our TNG50 simulation. There also is significant redshift evolution of SB$_0$ for the processed photons, roughly scaling with $(1+z)^4$. In the MUSE data, we find no significant correlation, neither with stellar mass nor redshift.}
\label{fig:rprofiles_SBcentral}
\end{figure}

In addition to the two LAH size measures $r_{1/2}$ and $r_0$ we also calculate the central surface brightness SB$_0$. Figure \ref{fig:rprofiles_SBcentral} shows the trend between SB$_0$ and galaxy stellar mass, from $z=2$ to $z=5$ (different line colors). We derive this value for both the intrinsic (dashed lines) and processed photons (solid lines). For the latter, we shade the central 68\% of outcomes in a given mass bin. Additionally, we show the results from the MUSE UDF data set (circles), colored to match the nearest simulated redshift for comparison.

First, we see a strong correlation of the peak LAH brightness with galaxy stellar mass. This trend was previously noted in Figure \ref{fig:rprofiles_stacked} for the simulated halos. Over four orders of magnitude in stellar mass, the central surface brightness value increases by roughly two orders of magnitude. The evolution of SB$_0$ as a function of redshift is minimal for intrinsic photons. While the intrinsic surface brightnesses are subject to cosmological dimming, this seems to be countered by the increased specific star formation rate for halos at higher redshifts. For processed photons, CGM diffusion and IGM attenuation suppress SB$_0$ with a scaling of roughly $(1+z)^4$.

In comparison to the clear correlation in the simulations, the MUSE data does not show such a strong relationship between central surface brightness and either stellar mass or redshift. Although the SB$_0$ values are in reasonable agreement where the bulk of the observed systems reside, $10^8 < M_\star/\rm{M}_\odot < 10^9$, the flat trend in the data leads to lower inferred values at higher stellar masses, when compared to those in TNG50. This is certainly caused in part by systematic uncertainties in the observational determinations of stellar mass, which are acknowledged for the observational estimates in~\cite{feltre20}. These uncertainties in part wash out the trend with stellar mass found here. Even more importantly, we speculate that this difference arises due to our omission of a model for unresolved dust attenuation and stochasticity on the smallest scales, as we discuss below.

\section{Discussion}
\label{sec:discussion}

\subsection{The Shape and Nature of Lyman-alpha Halos}

Based on our results we now discuss the implications for the shape and nature of Lyman-$\alpha$ halos. In particular, is there a typical shape for LAH radial profiles, is there a common cause for this shape, and what is the resulting interpretation of the observations? In Section \ref{sec:smallscales}, we focus on `small' scales, of order of $\sim 10\ $pkpc, where LAHs are detected around star-forming galaxies. In Section \ref{sec:largescales} we discuss larger scales, and profile flattening, as accessible in current and future intensity mapping studies.

\subsubsection{Ly$\alpha$ Profiles at Small Scales}
\label{sec:smallscales}

We have shown that the radial profiles of our simulated LAHs are primarily shaped through rescattered photons which originate from star-forming regions in the central galaxy of a halo on scales around $\sim 10\ $pkpc. This rescattering gives rise to a smoothing of the surface brightness maps and radial profiles that is larger than typical PSFs ($0.7\ $arcsec adopted herein). Beyond this distance median radial profiles tend to steepen rapidly before flattening on scales above $\sim 20\ $pkpc.

We find typical exponential scale lengths $r_0$ of $\sim 4\ $pkpc with little to no correlation with stellar mass. As radial profiles are largely dominated by emission from star-formation on these scales, the typical shape of rescattered photons from the central galaxy sets this typical $r_0$ and leads to the lack of correlation with stellar mass. If diffuse emission through collisions and recombinations were the dominant source of LAH photons, we would infer much larger scale radii $r_0$ and a strong correlation with stellar mass with scale lengths starting at $\sim 5\ $pkpc growing to $\sim 15\ $pkpc between $10^8$ and $10^{10}\ $M$_\odot$. Thus, $r_0$ and its mass dependency can serve as a discriminator between rescattered photons from star-forming regions and diffuse emission in observations. This holds even if the relative importance of the different emission sources is not precisely correct in our simulations.

An important modification of the \Lya emission stems from local ionizing sources. TNG's simplified on-the-fly treatment of radiation stemming from AGN is already reflected in our results. In particular, in our fiducial sample of LAH candidates with stellar masses of \smassdflt 1030 out of 1189 halos host a SMBH and incorporate their ionizing flux. While we expect that additional ionizing flux from stellar populations could additionally boost diffuse emission, we only find a small impact of local ionizing radiation from AGN on the radial profiles on small scales measured by the exponential scale radius $r_0$. As the AGN UV radiation (when present) dominates over that of the stellar population at lower redshifts considered, we similarly do not expect our $r_0$ results to significantly change if stellar radiation was incorporated.

Interestingly, as the scale length $r_0$ keeps growing toward higher masses for emission sourced by cooling and the UV background, high mass halos in TNG50 reach the lower end of extents observed in LABs. While a future dedicated study is required, this might strengthen the case of such diffuse emission sources (without local sources ionizing flux) causing observed LABs.

In comparison to data, our results for the stacked profiles are consistent with MUSE UDF observations presented in \cite{leclercq17}. At the level of individual LAH profiles, we similarly find good agreement, such that there are numerous analogs in TNG50 which have compatible \Lya radial profiles. This agreement is noteworthy as we have no tuned or calibrated parameters in our \Lya modeling. 

In the quantitative comparison of $r_0$ for individual LAHs we find up to a $20\%$ mismatch at $z=3$, which grows to a factor of two at $z=5$. Similarly, the central surface brightness values from \citet{leclercq17} show significant scatter and tend to be below those obtained from TNG50 at the highest stellar masses. Despite these regimes of tension, the observations show no clear mass or redshift evolution in either $r_0$ or SB$_0$ values, which together with the relatively small sizes implies that observed LAHs in the MUSE UDF sample are sourced by star-formation. 

Although a rigorous comparison between simulated and observed LAHs is complicated by several systematic uncertainties, the SB$_0$ and $r_0$ tensions hint at possible shortcomings of our \Lya modeling. In Figure \ref{fig:rprofiles_expR0} we found that considering star-formation alone gives the smallest exponential scale lengths. At $z=3$, we found $r_0=3.1_{-0.7}^{+1.4}\ $pkpc in this case, which is very similar to the MUSE UDF estimate of $3.1_{-0.6}^{+1.0}\ $pkpc. The simplest explanation is that we have underestimated the \Lya luminosity from star-forming regions, or overestimated the \Lya luminosity from other sources. Further, the strong correlation of SB$_0$ with stellar mass found in our models is not clearly present in the data. As we assume a fixed relation between star-formation and \Lya luminosity this outcome is not surprising, and can be alleviated by developing a more realistic model for the underlying relation between \Lya and SFR, as discussed in Section~\ref{sec:dustdiscussion}.

\subsubsection{Ly$\alpha$ Profiles at Large Scales}
\label{sec:largescales}

\begin{figure}
\centering
\includegraphics[width=1.0\linewidth]{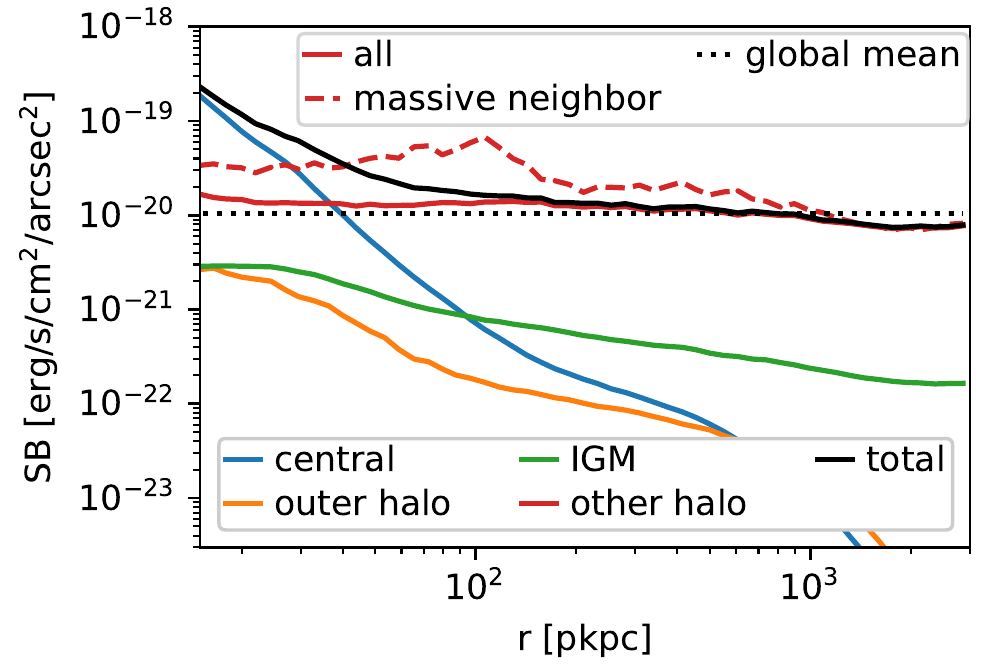}
\caption{Radial profiles for halos with \smassdflt{} at $z=3$ decomposed into emission origin and extending to large-scales of $3000\ $pkpc radii in TNG50. Here we neglect PSF effects and integrate photon contributions from within $\pm 600\ $pkpc of each emitter along the line of sight. The overall profile (solid black) strongly flattens out as the very flat contribution from other halos starts to dominate after $40\ $pkpc. Contributions from the outer halo (orange) and the IGM (green) are negligible in the stacked profiles, while nearby neighbors strongly boost the signal (solid red). To emphasize this environmental effect, we also show the subsample of 127 (out of 1189) halos that have a more massive halo, by at least a factor of 10, within 0.5 pMpc (dashed red). The black dotted line shows the surface brightness based on the global mean of the luminosity density in the simulation.}
\label{fig:veryextended}
\end{figure}

Although much of our analysis has focused on the inner regions of LAH profiles, $\lesssim 20\ $pkpc, we here consider the general shape of extended \Lya profiles where the impact of environment reveals itself.

The most prominent feature in the median stacked profiles of Figure \ref{fig:rprofiles_stacked}, regardless of stellar mass, is the significant `flattening' beyond the inner few 10s of pkpc. In Figure \ref{fig:veryextended} we take advantage of our global, large-volume RT calculation and present the radial profiles of halos with \smassdflt{} out to $3000\ $pkpc. Here we neglect the PSF, and integrate through $\pm 600\ $pkpc from the emitter's position along the line of sight, which roughly reflects the spectral resolution of the HETDEX survey. 

We decompose this median surface brightness profile based on photon origin (different colored lines). For reference, the expected surface brightness given the average luminosity density in TNG50 is shown with the black dotted line. We find that the contribution from other halos (solid red line) is nearly constant with distance and reaches down to roughly match the mean global. This `other halo' contribution starts to dominate the surface brightness profile beyond $100\ $pkpc by a factor of more than 10. To emphasize this environmental effect, we also show a stack of the subsample of halos that have a close ($\leq 0.5\ $pMpc) massive neighboring halo, by a factor of ten or more, as the red dashed line. For those galaxies there is an elevated plateau above the global mean that only starts to drop around $100\ $pkpc.

This behavior is a proximity effect whereby nearby halos give rise to an effectively elevated background. As a result, emission from star-formation from nearby neighbors dominates the observed radial profiles over diffuse emission from the outer halo and the IGM.
\cite{Zheng2011b} find a similar flattening effect from rescattered photons from other halos in their simulations, but are unable to assess the relative importance of diffuse emission, which remains untreated. In contrast, \cite{Lake2015} do incorporate both central sources and diffuse emission, reporting a flattening up to scales of $80\ $pkpc that in equal parts stems from the two emission sources, in contrast to our results where emission from other halos dominates over IGM emission.

We note that modeling of this proximity effect remains tricky: off-scattered photons in a targeted halo originating in a neighboring halo could potentially not be modeled by a classic 2-halo term as this would merely capture the overlap between different halos' profiles. Here -- in addition to the possible overlap of profiles -- we have \Lya photons freely traveling through the IGM to a neighboring halo at which point scatterings will trace out part of the targeted halo.

Our findings are qualitatively compatible with stacking results by~\cite{Matsuda2012} showing a strong correlation of the flattening in the overdense regions. 
\cite{wisotzki18} find significant flattening up to scales of around $\sim 50\ $pkpc, for which they consider UVB fluorescence as a potential source. In contrast, we find that towards large radii diffuse emission outside of halos contributes less than 10\% to the surface brightness (see Figure~\ref{fig:veryextended}), while we find that scattering of \Lya photons from galaxies dominate the flattening. 
More recent \Lya intensity mapping results at $z=5.7$ and $6.6$ by \cite{Kakuma2019} found reasonable agreement with the MUSE stacking \citep[see also][]{matthee20} and the simulated profiles of \cite{Zheng2011b} on small scales ($r<150\ $ckpc). However, they cannot confirm the proximity effect and flattening found here for larger scales. We note a further complication, that this flattening could be removed in large part or entirely in narrow-band surveys due to the required background subtraction. However, a careful examination of upcoming data, such as with HETDEX, could reveal this proximity effect, particularly by stacking based on the presence of massive neighbor number or environmental density.

\subsection{Current Limitations and Future Outlook}
\label{sec:limitationsoutlook}

\subsubsection{Model Limitations: TNG and its gas state}
\label{sec:dustdiscussion}

Our results depend inseparably on the outcome of the underlying cosmological simulation as well as our \Lya radiative transfer post-processing, inheriting limitations from each. The basis of our \Lya calculations is the TNG galaxy formation model \citep{weinberger17,pillepich18a} and the TNG50 simulation in particular. Although the TNG model has been shown to reproduce a large variety of galaxy and CGM properties in broad agreement with observations (see Section \ref{sec:sims}), 
it does not treat local ionizing radiation self-consistently as done in radiation-hydrodynamical (RHD) simulations, that are now becoming increasingly feasible for individual zoom simulations \citep[e.g.][]{Rosdahl2012,mitchell21} as well as cosmological volumes down to redshifts $z=5-6$~\citep{Gnedin2017,Rosdahl2018,Ocvirk2020}.
As a result, our knowledge of the gas ionization and temperature state is limited to the outcome of the physical modeling in our simulation, as is true in all galaxy formation simulations. As TNG50 remains unfeasible as RHD simulation, it substitutes such approach with an on-the-fly treatment of AGN as the dominant ionizing radiation source as studied in Appendix~\ref{sec:ionizingsources}. Towards higher redshifts, the relative importance of stellar ionizing sources increases, thus potentially hinting at larger uncertainties of TNG's model only incorporating AGN's radiative feedback.

While particularly emission from collisional excitation strongly depends on the temperature~\citep{Furlanetto2005, Faucher2010} and subsequently can boost emission, we found that the diffuse emission remains subdominant to scattering from the central sources even if radiation from AGNs is present. If central \Lya emission from SMBHs themselves is incorporated, this finding will be further strengthened. Predicted central surface brightnesses when only considering collisional excitations from the CGM in TNG, even with AGN radiative feedback, are commonly an order of magnitude too small compared to the MUSE observations. Apart from this normalization problem, collisional excitations from TNG's predicted temperature and ionization state show scale lengths significantly too extended for compatibility with observations (see Figure~\ref{fig:rprofiles_expR0}). Reconciliation with observations in case collisional excitations are the dominant source of LAHs thus requires significantly different ionization and temperature radial profiles from those found in TNG. 
Independent constraints on the density, ionization state and temperature from other observational probes would be useful to assess TNG's model and check resulting conclusions for predicted LAHs. In addition, these findings should be revisited with future RHD simulations that come closer TNG's sample size and studied redshift range.

In addition, the TNG50 simulation has a resolution of $\sim 100$\,pc in the dense interstellar medium, which is furthermore modeled with an effective two-phase model. As above, this implies that a number of approximations and simplifications must be adopted in both the emission and transport of \Lya (see Section \ref{sec:emission}). Importantly, in the present work, we have adopted a simple mapping from the star-formation rate of a gas cell to its \Lya emissivity. We have also neglected dust in its entirety, as well as sub-resolution (unresolved) density structure during the radiative transfer step.

\subsubsection{Model Limitations: ISM emissivity}
\label{sec:SFluminosity}

As discussed, we do not include the destruction of \Lya photons by dust in our \Lya radiative transfer (although we have investigated its impact, see Appendix~\ref{sec:dust}). Such destruction would primarily take place in the ISM, where resolution is marginal for capturing the complex density and ionization structure relevant for the \Lya radiative transfer.
If however we do not explicitly treat dust, we potentially overestimate the SF luminosities in Eqn.~\eqref{eq:lumSF}, where we do not explicitly model dust destruction. We review the emission model for stellar populations here, as such an overestimation would strongly affect our conclusions of central \Lya emission dominating observed LAH profiles through CGM scatterings.

Equation~\eqref{eq:lumSF} is derived by integrating the ionizing flux from stellar population synthesis models, and a conversion of $\sim 2/3$ of ionizing photons into \Lya \citep{Dijkstra2019}. The proportionality factor between SFR and \Lya emission depends on a range of factors such as the stellar population synthesis model, the escape of ionizing flux, the stellar population's age and metallicity, and the initial mass function (IMF)~\citep{Furlanetto2005}. Overall, this relation should only be seen as a rough estimate with an uncertainty of a factor of a few. Furthermore, observations, such as from H$\alpha$, Ly$\alpha$/$H\alpha$ and Ly$\alpha$/H$\beta$ show a large scatter in the relation of hydrogen line emission to SFR \citep{Kennicutt1998,Blanc11,weiss21}.

Given the large modeling uncertainties shown, we adopted this SFR-\Lya relation as a rough estimate, which could potentially be rescaled after comparison with observations. However, as we show in the following, such a calibration is not necessary for the mass range \smassdflt{}, where we already obtain reasonable agreement with observations. This also implies that, while not used in the derivation of the relation, \Lya dust extinction is nevertheless effectively captured for the \smassdflt{}. We note however, that at the high mass end luminosities can be significantly overestimated, and the role of dust becomes critical.

In Figure~\ref{fig:rprofiles_SBcentral}, we show the central luminosities as predicted in our simulations compared to MUSE observations in~\cite{leclercq17}. We find that the simulations show similar or slightly lower central surface brightnesses when compared with the observational data. 

We have also compared the LAH luminosity function (LF) of our samples with observational LAE luminosity functions at $z=2.2$, $3.1$ and $5.7$ \citep[respectively based on][]{Konno2016,Ouchi2008,Konno2018} (not shown). 
We calculate LAH luminosities from the photon contributions of the targeted halos falling within a 3 arcsecond aperture. We find very reasonable agreement of the observations with our intrinsic luminosity functions in the luminosity range $\sim 3\times 10^{41}\ $erg/s to $\sim 2\times 10^{42}\ $erg/s, which includes at least 68\% of all halos with stellar masses \smassdflt{} for each redshift $z=2$ to $z=5$ in TNG50. Quantitatively comparing to~\cite{Konno2016}, we find that at $L=1.4\times 10^{42}$ erg/s, our intrinsic emission in fact underestimates the LF by just 0.16 (0.32, 0.17, 0.06) dex for $z=2$ (3,4,5). For the processed photons, our LFs underestimate the observed LFs by 0.19 (0.10, -0.11, -0.63) dex at $z=2$ (3,4,5). That is, our preliminary look already shows promising agreement, while a detailed analysis of the LFs remains a topic for future work.

We are thus confident that, on average, for the fiducial stellar mass range of \smassdflt, the luminosities reasonably match observations. We thus adopt the simple SFR-\Lya relation without readjustment for dust attenuation. However, we overestimate the high luminosity end of the LF, and thus inaccurately capture the high-mass LAHs' shape. This might also affect the overall amplitude of the flattening from neighboring galaxies for the fiducial stellar mass range.

In observations there is a large scatter between the star-formation rate inferred from \Lya{} luminosity compared to UV based estimates \citep{Santos2020,Runnholm20}. Typically, the median SFR estimate from \Lya{} exceeds that from UV measurements below $\sim 10\ $M$_{\odot}$/yr, but decreases above this value. This systematic trend implies a larger suppression of \Lya{} escape for emitters with higher SFR and stellar mass. In addition, there is non-negligible scatter between individual objects with similar properties. It is thought that \Lya{} emission traces the most recent star-formation \citep{Santos2020}, which could be highly time variable, and is modified by the complex small-scale neutral hydrogen distribution and kinematics \citep{Blanc11}.

In our simulations we have instead assumed a strict proportionality between \Lya luminosity and SFR at the resolution scale, as given by Equation~\eqref{eq:lumSF}. We find that subsequent scattering only adds minor scatter to this correlation, implying that sub-resolution stochasticity may be required. As discussed in Section \ref{sec:smallscales}, we tend to slightly overestimate the $r_0$ metric of LAH sizes in comparison to the MUSE UDF sample, and generally find strong trends of both $r_0$ and SB$_0$ with stellar mass, which are less clear in the data. We suspect that adopting an observationally motivated scatter between \Lya luminosity and UV inferred SFR would also alleviate the tensions noted in the central surface brightness SB$_0$ trends between our simulations and the data. We ran a simple test to test the impact of including a naive dust treatment (Appendix \ref{sec:dust}), which demonstrates that the central brightness SB$_0$ is increasingly attenuated towards higher stellar masses, as dust counters the generally higher SFRs and thus \Lya luminosities of those halos.

While subgrid modeling and dust, along with intrinsic scatter between \Lya{} luminosity and SFR, should help reconcile the central surface brightness SB$_0$ comparison in Figure~\ref{fig:rprofiles_SBcentral}, we would additionally require the mean \Lya{} luminosity from star-formation to increase in order to obtain smaller LAH sizes. For the typical objects considered as LAH candidates with \smassdflt{} here, \citet{Santos2020} find SFR$_{\mathrm{Ly}\alpha}$/SFR$_\mathrm{UV}\sim 2$ for SFR $<10\ $M$_\odot$/yr with significant scatter. Beside the simplistic model assumptions with significant uncertainties as previously discussed, we would need to incorporate \Lya emission from AGN in future work, which would boost emission and naturally induce scatter in the SFR-\Lya relation.

In conclusion, we demonstrated that the current ISM emissivities appear reasonable given the simplistic model, but future work on the effective modeling of dust and a modified SFR-\Lya relation (reflecting varying ISM environments and potential AGN presence) in both proportionality factor and its scatter might help to simultaneously diminish tensions of SB$_0$ and $r_0$ in future work.

\subsubsection{Future Directions}

An important improvement for future \Lya radiative transfer modeling in large-scale cosmological simulations such as TNG50 will be a treatment of complexities below the resolution scale ($\sim 100\ $pc). Such a model could either implement explicit numerical subgrid models \citep{Hansen2006,Gronke2017b} and retain our current principle of a parameter-untuned model. Alternatively, we could adopt an effective parameter-based model, incorporating observational findings to motivate \Lya production and escape fractions from star-forming regions \citep{weiss21}.

In future work we will also use our coupling of \textsc{voroILTIS} and TNG50 to examine two key areas: environmental imprints on large-scale observations, and the information content of the spectral dimension. We found a significant redistribution of photons from star-forming regions to large scales ($\gtrsim 100\ $ kpc) which might affect the interpretation of \Lya{} intensity mapping experiments. Furthermore, spectral modeling of \Lya emission in cosmological volumes remains challenging given its multi-scale nature \citep{byrohl20a,song20}. However, detailed and spatially resolved spectral information on \Lya-halos is increasingly becoming available \citep{claeyssens19,leclercq20}, and promises to offer significant insight into the kinematics and small-scale structure of hydrogen gas in the CGM of dark matter halos across cosmic time.

Finally, we have here focused on lower mass star-forming galaxies with $M_{\rm halo} < 10^{12}$\msun. Above this threshold, the AGN in more massive galaxies are known to have a significant impact on the ionization state of the CGM and the \Lya scattering processes occurring therein. Our current radiative transfer methodology does not account for local radiation fields from AGN, but this is a natural extension of \textsc{voroILTIS} which will allow us to compare to the many rich observational data sets of \Lya{} emission around quasars.

\section{Conclusions}
\label{sec:conclusions}

In this paper we develop a technique to perform full radiative transfer calculations to trace resonantly scattered \Lya emission, and the Ly$\alpha$ halos (LAHs) around galaxies, at $2 < z < 5$. We do so by post-processing the TNG50 cosmological magnetohydrodynamical simulation \citep{nelson19b,pillepich19} of the IllustrisTNG project.

This large volume offers a powerful statistical sample of thousands of LAHs across an unprecedented halo mass range and across diverse environments, together with a high resolution of order 100 physical parcsecs in the dense interstellar medium. Furthermore, our new radiative transfer code \textsc{voroILTIS} (\textcolor{blue}{Byrohl et al., in prep}) operates natively on the global Voronoi tessellation of the TNG simulation volume, incorporates both diffuse and galaxy emission, and self-consistently accounts for attenuation within the intergalactic medium.

This allows us to carry out a detailed investigation into the origins, physical properties, and emission sources that shape Lyman-$\alpha$ halos. At the same time, the realism of the underlying TNG50 simulation enables us to make quantitative connections between LAH and galaxy properties. Our main findings are:

\begin{itemize}
    \item Star-forming galaxies with $10^7 < M_\star / \rm{M}_\odot < 10^{10.5}$ at $2 < z < 5$ emitting \Lya photons are surrounded by extended Lyman-alpha halos (LAHs). We present the stacked, median predictions for TNG50 LAHs as a function of galaxy mass and redshift. The radial surface brightness profiles of LAHs have a characteristic shape comprised of a rapid, exponential decline followed by a large-distance flattening. This flattening arises from the density structure probed by rescattering photons as well as in-situ diffuse emission.
    \item Scattered photons from star-forming regions are the dominant contributor to LAH profiles on typically observed scales $r\lesssim 20\ $pkpc. Given the importance of scattered photons, we stress the need to use radiative transfer simulations or semi-analytic expressions capturing such behavior. At larger distances, contributions from diffuse emission via recombinations and de-excitations become equally important. 
    \item On larger scales $r \gtrsim 30\ $pkpc we find that the flattening of LAH profiles is actually dominated by rescattered photons that originate from other nearby halos, rather than the primary halo itself. This proximity effect is boosted in high density environments, and should be observable. A careful reproduction of a survey's background subtraction and wavelength window are needed for a detailed comparison of this flattening to observations.
    \item Characterizing LAH sizes, we find that their half-light radii $r_{1/2}$ are of order $5\ $pkpc at $z=2$, increasing to $\sim 15\ $pkpc at $z=5$. This signposts a significant redistribution of photons due to higher neutral hydrogen densities at higher redshifts. The exponential scale lengths $r_0$ also increase slightly towards higher redshift for galaxy stellar masses of \smassdflt{}. Neither $r_{1/2}$ and $r_0$ show clear trends with mass. In contrast, our fiducial model shows a strong positive correlation of the central surface brightness SB$_0$ with both stellar mass and redshift.
    \item While AGN radiative feedback adds significant heating and ionization to the surrounding CGM, subsequently boosting intrinsic emission from both collisional excitations and recombinations, we only find a marginal impact on the emission mechanisms' relative importance and the overall scale lengths $r_0$ for LAHs with AGN activity in TNG.
    \item We compare the stacked, median LAH radial profile between TNG50 and data for galaxies with \smassdflt{} from the MUSE UDF and find good qualitative agreement. We also demonstrate, by finding statistically consistent simulated analogs for individual observed profiles, that the simulation successfully reproduces the diversity of observed LAHs.
    \item For the quantitative comparison to observational results of LAH sizes as measured by half-light radii $r_{1/2}$ and exponential scale lengths $r_0$, we find agreement at the level of $\sim 20\%$ at $z = 3$, with the $r_0 \sim 4\ $pkpc of our simulated profiles slightly above those of the MUSE UDF data set ($r_0\sim 3\ $pkpc). This difference increases with redshift, to a factor of two at $z = 5$. Similarly, we find that the central surface brightness SB$_0$ of LAHs is in good agreement at low stellar mass, but such agreement becomes progressively worse towards higher $M_\star$. Both tensions arise because LAH properties tend to correlate strongly with both galaxy mass and redshift in the simulations, but less so in the data. Barring important differences in the property distributions of the selected galaxies between the simulated and observed samples, we attribute these correlations to a number of simplifying assumptions in our \Lya modeling, and discuss future improvements.
\end{itemize}

The extended \Lya emission around galaxies and quasars at $z>2$ provides an insightful window into many aspects of galaxy formation and evolution. Here we have demonstrated the power of coupling a large-volume cosmological hydrodynamical simulation with a global \Lya radiative transfer modeling. Future improvements in our treatment of unresolved small-scale gas structure and local radiation fields from AGN and stellar populations will enable interpretation of additional datasets and upcoming surveys, from large-scale intensity mapping experiments to highly detailed, spectral data from targeted IFU surveys.

\section*{Data Availability}

Data directly related to this publication and its figures is available on request from the corresponding author. The IllustrisTNG simulations, including TNG50, are publicly available and accessible at \url{www.tng-project.org/data} \citep{nelson19a}.

\section*{Acknowledgements}

We thank Floriane Leclercq for providing the full Ly$\alpha$ SB(r) data shown in \citet{leclercq17}. We thank Anna Feltre, Roland Bacon, and the MUSE UDF team for providing SED fitting results prior to their release. We thank Max Gronke for fruitful discussions on \Lya radiative transfer. DN acknowledges funding from the Deutsche Forschungsgemeinschaft (DFG) through an Emmy Noether Research Group (grant number NE 2441/1-1).

The primary TNG50 simulation was carried out with compute time granted by the Gauss Centre for Supercomputing (GCS) under GCS Large-Scale Project GCS-DWAR (PIs: Nelson and Pillepich) on the GCS share of the supercomputer Hazel Hen at the High Performance Computing Center Stuttgart (HLRS). GCS is the alliance of the three national supercomputing centres HLRS (Universit{\"a}t Stuttgart), JSC (Forschungszentrum J{\"u}lich), and LRZ (Bayerische Akademie der Wissenschaften), funded by the German Federal Ministry of Education and Research (BMBF) and the German State Ministries for Research of Baden-W{\"u}rttemberg (MWK), Bayern (StMWFK) and Nordrhein-Westfalen (MIWF). Additional simulations and analysis were carried out on the supercomputers at the Max Planck Computing and Data Facility (MPCDF). 
MV acknowledges support through NASA ATP grants 16-ATP16-0167, 19-ATP19-0019, 19-ATP19-0020, 19-ATP19-0167, and NSF grants AST-1814053, AST-1814259,  AST-1909831 and AST-2007355.

\bibliographystyle{mnras}
\bibliography{references}

\appendix

\section{Modeling Assumptions}
\label{sec:modelassumptions}

In this appendix, we assess various modeling assumptions along with numerical convergence of our results. Unless stated otherwise we focus on the star-formation emission source only, and derive radial profiles neglecting rescattered light from other halos, as we limit ourselves to rerunning the radiative transfer on halos with stellar masses \smassdflt{}.

\subsection{Hydrodynamic Resolution}
\label{sec:hydrores}

\begin{figure}
\centering
  \includegraphics[width=1.0\linewidth]{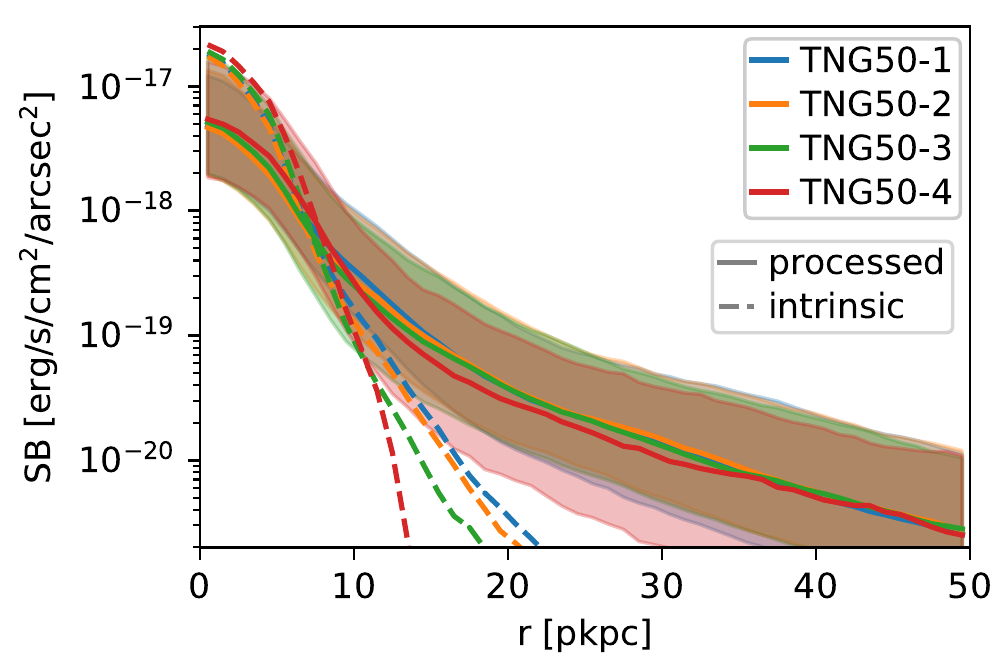}
  \caption{Median stacked Ly$\alpha$ surface brightness radial profiles at $z=3$ for a fixed stellar mass \smassdflt at different resolution runs of TNG50. TNG50-1 has a mass resolution of  $m_\mathrm{baryon}=8.5 \times 10^4\ $M$_\odot$, which increases by a factor of eight for each lower resolution run. Only star-formation is considered as an emission source, and scattered light contributions from other halos are ignored. The dashed lines correspond to the intrinsic emission radial profiles with a corresponding solid line for the profiles after radiative transfer. The shaded regions show the 16th to 84th percentiles. Differences between resolution levels are generally minor.}
\label{fig:rprofiles_resolution}
\end{figure}

In Figure~\ref{fig:rprofiles_resolution} we show the Ly$\alpha$ surface brightness radial profiles at different hydrodynamical resolutions of TNG50 runs stacked at fixed stellar mass \smassdflt. Overall, we find the stacked profiles to be very robust over the different resolution levels shown. For fixed stellar bins, we find the intrinsic emission to slightly expand with increasing resolution. For the processed photons, there is a slight flattening of the stacked profiles between $10$ and $20\ $pkpc.

Despite this the inferred $r_0$ values are effectively invariant with changing resolution: we find $3.1_{-0.6}^{+1.4}\ $pkpc (TNG50-1), $3.0_{-0.7}^{+1.3}\ $pkpc (TNG50-2), $2.7_{-0.5}^{+1.2}\ $pkpc (TNG50-3) and $3.0_{-0.5}^{+0.8}\ $pkpc (TNG50-4). However, there appears to be a decrease in the scatter, which may reflect the simpler density structure at lower resolution. 

The resolution dependency appears to be significantly smaller than that in \cite{Zheng2011b}. We note that even the lowest resolution run presented here has a higher resolution than that study in the proximity of halos.

\subsection{Photon Package Count}
\label{sec:photonconvergence}

\begin{figure}
\centering
  \includegraphics[width=1.0\linewidth]{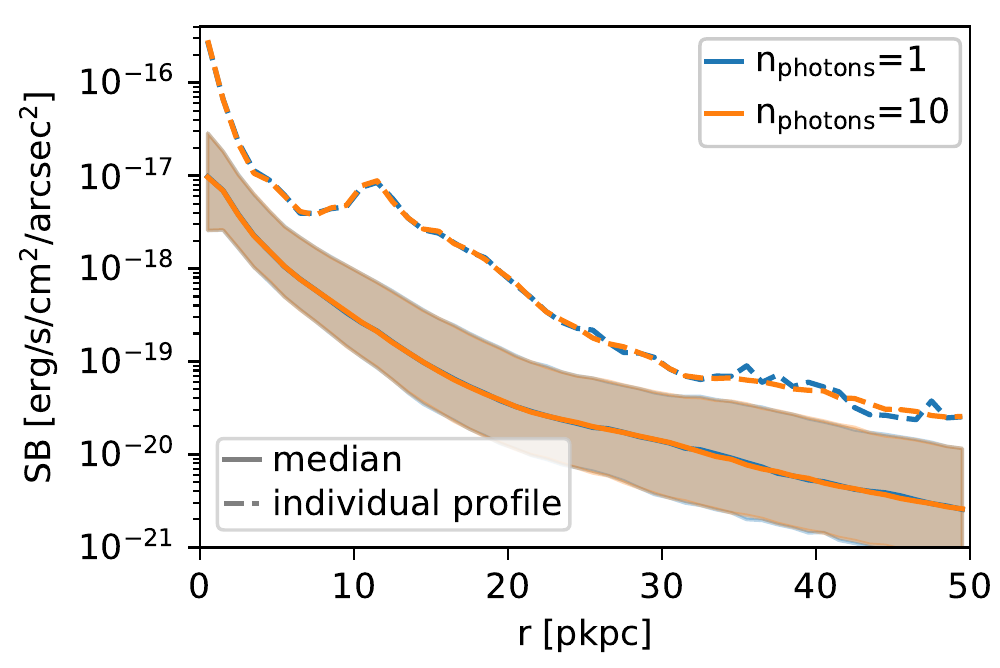}
  \caption{Median stacked Ly$\alpha$ surface brightness radial profiles (solid lines) at $z=3$ in TNG50 for a fixed stellar mass \smassdflt{} for $1$ (fiducial, blue) and $10$ Monte Carlo photon packages per star-forming cell (orange). To show the largest possible discrepancy, we also include the least-converged individual radial profile (dashed lines), which also demonstrates near perfect convergence in $n_{\rm count}$. Our results are clearly insensitive to this parameter.}
\label{fig:rprofiles_nphotons}
\end{figure}

In our fiducial runs, we spawn one photon package per Voronoi cell and emission source (star-forming region). Particularly for photons originating from central star-forming regions, there might be a convergence issue given the large volumes that are to be traced out at larger radii. We thus focus on convergence checks for the photon package count from star-forming regions. 

In Figure \ref{fig:rprofiles_nphotons} we show the radial profiles with varying $n_{\rm count}$ of initial photon packages. The median stacked profiles (solid) are already fully converged at our fiducial choice of $n_\mathrm{photons}=1$. As the most stringent test, we also plot the radial profile of the single halo with the largest sum of squared errors between both these two test runs -- the least converged individual profile. In this case we also find only minuscule deviations confirming that our choice of the photon count leads to a well converged result. 

We note that for halos with stellar masses between \smassdflt, our fiducial setup spawns $\sim 35000$ intrinsic photons: the intrinsically high resolution of TNG50-1 gives us already sufficient sampling.

\subsection{Input Spectrum}
\label{sec:inputspectrum}

\begin{figure}
\centering
  \includegraphics[width=1.0\linewidth]{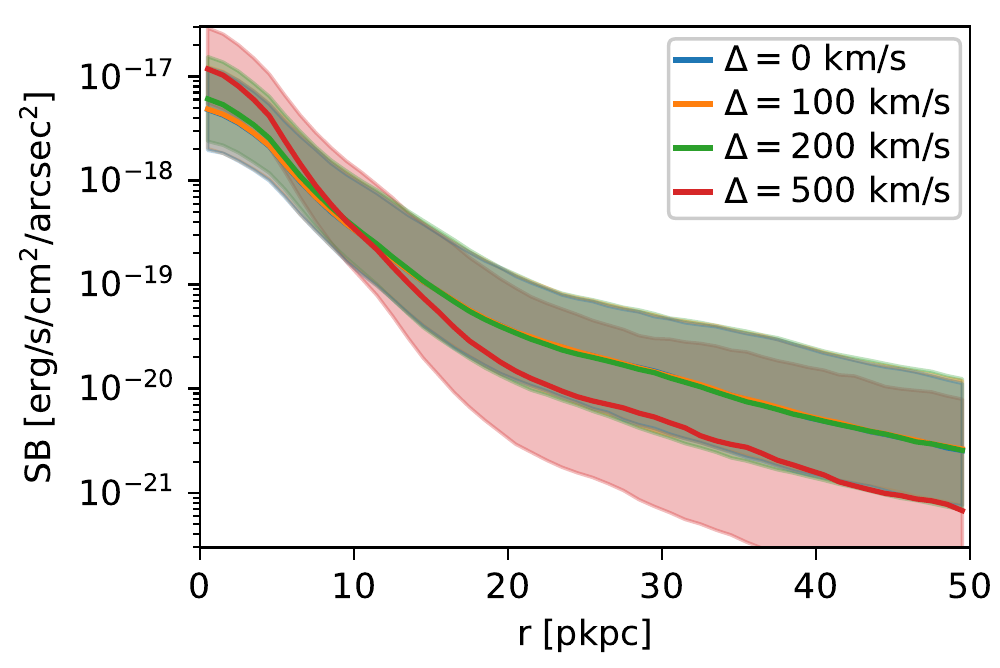}
  \caption{Median stacked Ly$\alpha$ surface brightness radial profiles (solid lines) at $z=3$ in TNG50 for a fixed halo mass bin of  $\log\left(\rm{M}_h/\rm{M}_\odot\right)\in\left[10.5,11.0\right]$. Here we contrast four different input photon wavelength offsets. In our fiducial model, photons are injected at the line-center within the rest-frame of the respective star-forming cell, indicated here as an offset of $\Delta = 0$ km/s from the line center (blue). Three non-zero offsets of $\Delta=100\ $km/s (orange), $\Delta=200\ $km/s (green) and $\Delta=500\ $km/s (red) are shown, where the shaded regions enclose the central $68\ $\% of SB(r) at fixed r. We find little difference in the resultant LAH profiles between the different input spectra choices ($\Delta$), except for the unrealistically large case of $500\ $km/s.}
\label{fig:rprofiles_sfroffset}
\end{figure}

In our fiducial model, photons are injected at the Ly$\alpha$ line-center in the rest-frame of the respective hydrodynamic cell. For diffuse emission, Equations~\eqref{eq:lumrec}/\eqref{eq:lumexc}, this is a fair assumption given that optical depths are moderate, although subgrid clumpiness would add complexity. However, in dense star-forming cells radiative transfer is significantly more complex due to small-scale dust, clumpiness, and ionization. Our radiative transfer simulations do not self-consistently capture these details. While attenuation on these scales only changes the overall radial profile normalization of the attenuated component, the spectral shape of emitted photons might have an effect on the radial profile shape itself.

\begin{figure}
\centering
  \includegraphics[width=1.0\linewidth]{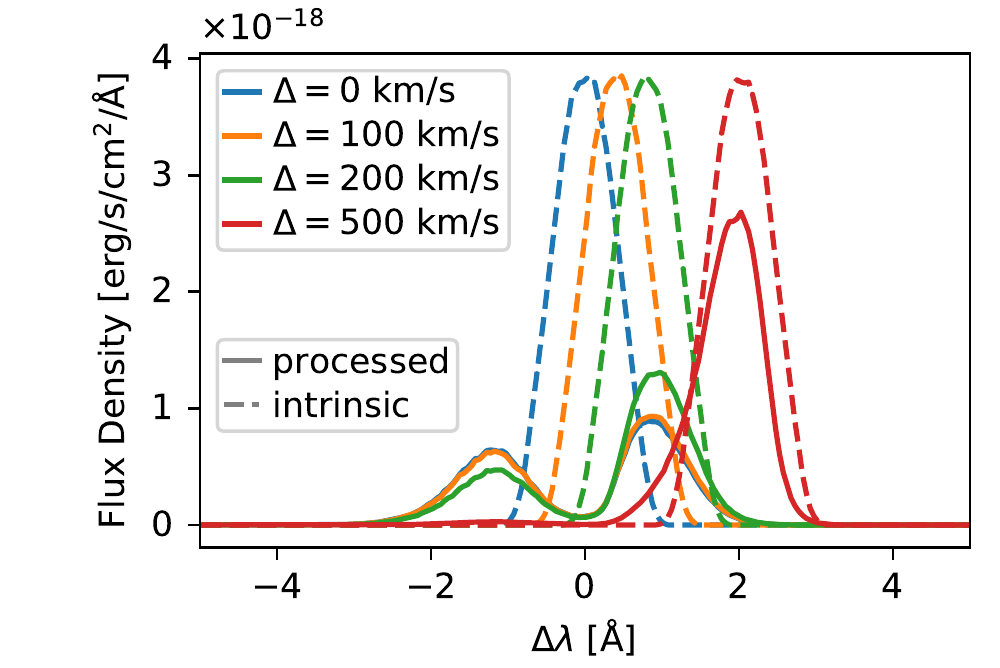}
  \caption{The mean stacked \Lya flux spectra at $z=3$ for a fixed halo mass $\log\left(\rm{M}_h/\rm{M}_\odot\right)\in\left[10.5,11.0\right]$ in TNG50 for a varied injected photon wavelength offset $\Delta\lambda$ from the \Lya line-center, integrating within a 3 arcsecond radius aperture. Dashed lines show the intrinsic photon frequency distributions, while solid lines show the emergent (processed) spectra. This corresponds to the same sample for which we show the stacked profiles in Figure \ref{fig:rprofiles_sfroffset}.}
\label{fig:spectra_sfroffset}
\end{figure}

As shown in similar cosmological simulations by \citet{Byrohl2019}, Ly$\alpha$ spectra often appear to have too much flux at wavelength larger than the line-center (too `blue') compared to observations. We thus consider an additional redshift for the injected photons, which naturally leads to a more realistic `red' spectrum. In Figure \ref{fig:rprofiles_sfroffset} we show the results of radiative transfer simulations where the redshift $\Delta$ from the line center has been varied between $0\ $km/s (fiducial model) and $500\ $km/s, considering emission from star-formation only. The corresponding intrinsic (dashed) and emergent (solid) \Lya spectra are shown in Figure \ref{fig:spectra_sfroffset}.

We find that the choice of $\Delta$ has virtually no impact, except for the unrealistically high offset of $500\ $km/s. We recover sizes of $3.1_{-0.7}^{+1.4}\ $pkpc ($\Delta=0\ $km/s), $3.1_{-0.7}^{+1.4}\ $pkpc ($\Delta=100\ $km/s), $3.0_{-0.7}^{+1.4}\ $pkpc ($\Delta=200\ $km/s) and $2.5_{-0.4}^{+1.0}\ $pkpc ($\Delta=500\ $km/s) for halos with \smassdflt. For large $\Delta$ the scale length $r_0$ decreases as less scatterings occur given the lower cross-section in the \Lya{} line profile wings, while interaction with the IGM also decreases as this is mostly driven by blue photons. Although the frequency distribution at injection will be important for future studies, it is not crucial for our study of LAH profiles.

\subsection{Dust}
\label{sec:dust}

In our fiducial model we have neglected dust. Comparison of luminosities with observations in Section~\ref{sec:SFluminosity} indicate that they are reasonable for the fiducial stellar mass range \smassdflt{} without further explicit dust modeling, although this will be necessary for the high mass end.
Dust modeling can either take the form of an additional attenuation factor of \Lya emission, particularly of star-forming gas cells, or an explicit dust treatment in the \Lya RT. 
In this section, we explore the option of including an explicit dust treatment in the \Lya RT, as a preliminary study. We neglect clumpiness, such that dust is smooth at the resolution scale of the simulation. We use the model for Milky Way like dust \citep{Weingartner2001} as implemented in \citet{Behrens2019}, which relies on the metallicity field in TNG50.

We expect that dust will primarily modulate the overall normalization of the radial profiles for the emission from the ISM given in Equation~\eqref{eq:lumSF}. Nevertheless, this can boost the relative importance of other emission mechanisms, and change the overall radial profiles. As dust content is related to gas-phase metallicity and galaxy mass, the impact of dust could alter the trends of LAH properties with mass, and be particularly important at higher galaxy masses.

\begin{figure}
\centering
  \includegraphics[width=1.0\linewidth]{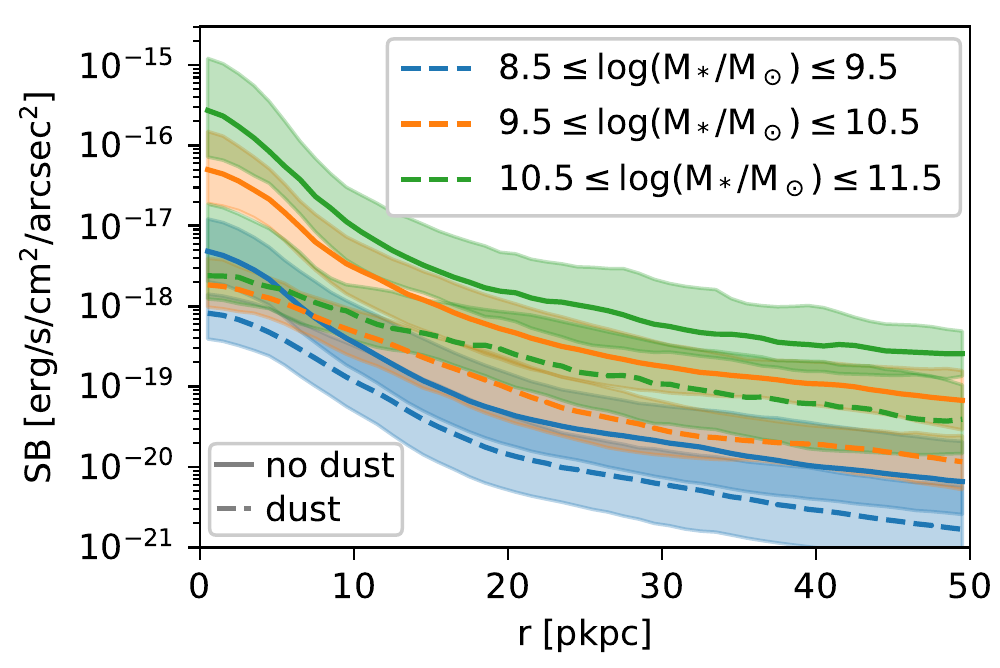}
  \caption{Median stacked Ly$\alpha$ surface brightness radial profiles in various stellar mass bins at $z=3$ in TNG50. In this plot, we only consider contributions from star-forming regions, where the impact from dust is most severe. We show the results with (dashed lines) and without (solid lines) dust modeling. We generally find a significant suppression of flux that significantly increases for higher stellar mass halos.}
\label{fig:rprofiles_dust}
\end{figure}

In Figure \ref{fig:rprofiles_dust} we show the impact of dust in our simulations at $z=3$ for emission from star-formation only. We contrast LAH radial profiles between the fiducial dust-free case (solid lines) and the dust included model (dashed lines). Our findings on the impact of dust are similar to those in \citet{Laursen2009}. In particular, surface brightness is increasingly suppressed in overdense, dusty regions. In addition, regions of lower density are uniformly suppressed as dust limits the escaping (and then rescattering) contributions from star-forming regions.

Dust attenuation strongly scales with the stellar mass. For example, the median attenuation for the central surface brightnesses is less than one order of magnitude for \smassdflt{}, but roughly two orders of magnitude for \smassdflthigh{}.

The inner $\lesssim 10\ $pkpc are particularly suppressed, decreasing the central surface brightness values and thus increasing the exponential scale lengths. For the scale lengths of halos with \smassdflt, we find an average $r_0 \simeq 3.1_{-0.7}^{+1.4}\ $pkpc without dust and $5.1_{-1.2}^{+2.4}\ $pkpc with dust. Both the median scale length and its variance within the sample increase with dust. The increase in scale length with this dust modeling is even larger when incorporating other emission sources, as they become relatively more important (see Section \ref{sec:results}). It is clear that future models will need to include at least a basic dust model.

\subsection{Impact of Local Ionizing Sources}
\label{sec:ionizingsources}

Properly accounting for the impact of local ionizing sources on the temperature and ionization state of the CGM requires, ultimately, full radiation-hydrodynamical simulations. However, this remains computationally infeasible for the present combination of galaxy sample size, resolution and redshift.

In this section, we assess the impact of local ionizing radiation from AGN and stellar populations. An on-the-fly treatment (i.e. self-consistently considering local sources' impact on cooling in each time step) is implemented for AGN radiative feedback in TNG. Hence, we can measure such ionizing sources' impact on the LAHs by comparing halo samples with and without SMBHs, as in Section~\ref{sec:AGN}. In the subsequent Section~\ref{sec:stars} we then discuss what impact ionizing radiation from stellar populations would have, given that this is not included in our models.

\subsubsection{Active Galactic Nuclei}
\label{sec:AGN}

\begin{figure*}
\centering
\includegraphics[width=0.48\linewidth]{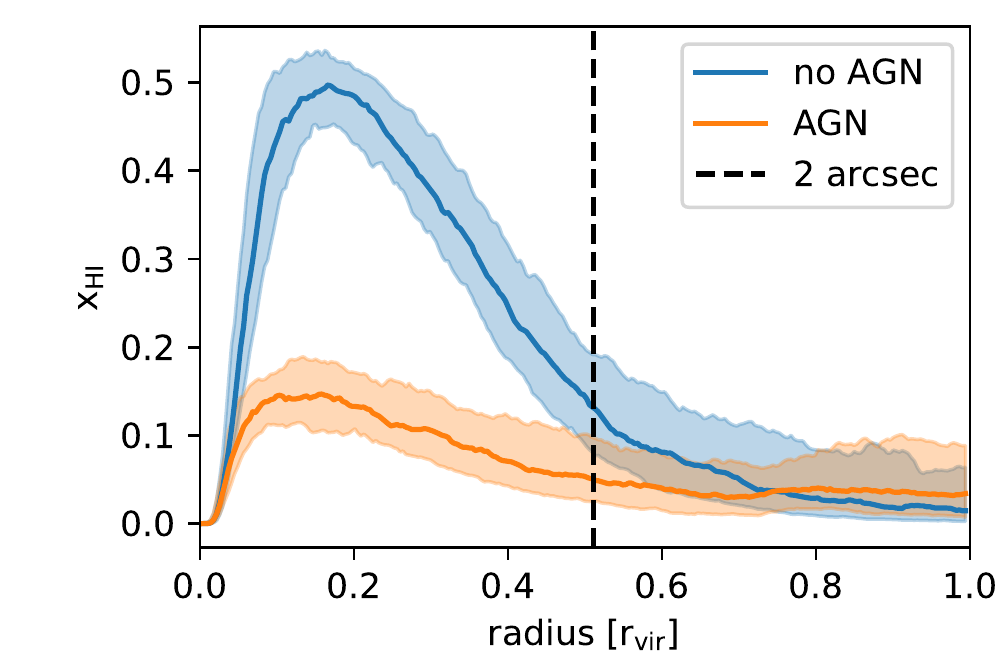}
\includegraphics[width=0.48\linewidth]{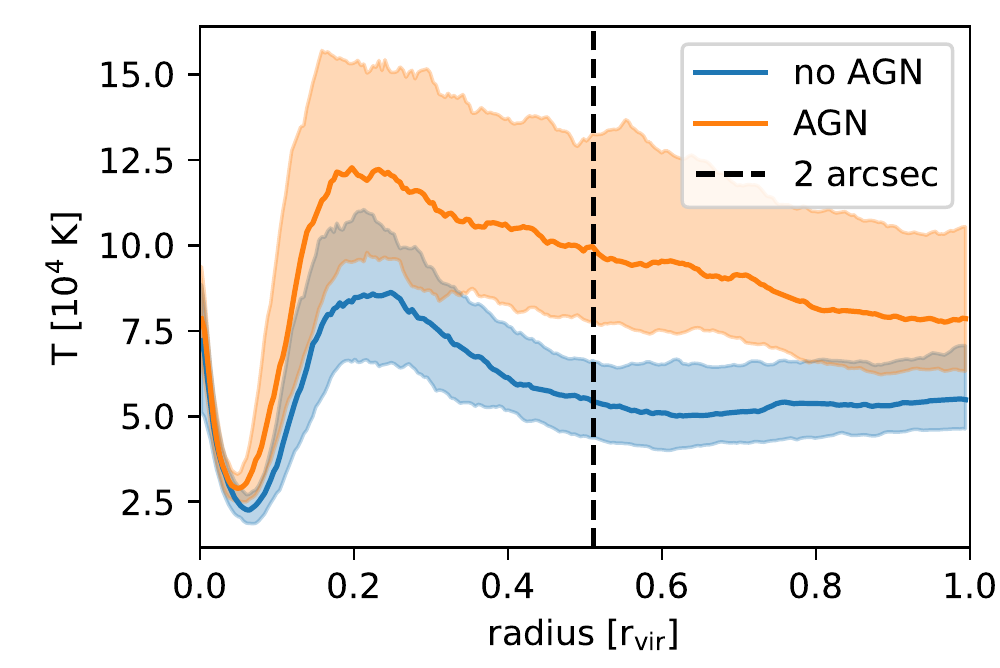}
\caption{Median radial profile at z=3.0 for \smassmini split into halos with/without AGN in TNG50 (148/122 halos) for the neutral hydrogen fraction (\textbf{left}) and temperature (\textbf{right}). Angular averaging uses mass-weighting. The shaded regions show the 16th to 84th percentiles. The vertical dashed line shows, in the median, the radius corresponding to 2 arcseconds.}
\label{fig:rprofiles_AGN_hydrostate}
\end{figure*}

Ionizing radiation from SMBH is modelled in TNG, which we will describe in the following. Note that we use the terms AGN and SMBH interchangeably here.

The intrinsic luminosity from black hole mass accretion at a rate $\dot{M}_\mathrm{BH}$ in TNG is taken to be
\begin{align}
L^{\mathrm{SMBH}}_\mathrm{bol} = (1-\epsilon_f)\ \tilde{\epsilon}_r\ \dot{M}_\mathrm{BH}\ c^2.
\end{align}
according to the radiative efficiency $\tilde{\epsilon_{r}}$ and feedback energy fraction $\epsilon_{f}$.
The ionizing luminosity escaping the galaxy is then given by
\begin{align}
\label{eq:L_UV_AGN}
    L^{\mathrm{SMBH}}_\mathrm{UV,esc} 
    = A\  f^{\mathrm{AGN}}_{\mathrm{esc}}\ L^{\mathrm{SMBH}}_\mathrm{bol},
\end{align}
where $A$ describes the bolometric correction factor of ionizing luminosity for the assumed spectral energy distribution and $f^{\mathrm{AGN}}_{\mathrm{esc}}$ incorporates the escape of ionizing flux from the galaxy as
\begin{align}
  f^{\mathrm{AGN}}_{\mathrm{esc}}=\omega_1\left(\frac{L^{\mathrm{SMBH}}_\mathrm{bol}}{10^{46}\ \mathrm{erg/s}}\right)^{\omega_2}
\end{align}
with $\omega_1=0.3$, $\omega_2=0.07$~\citep{Hopkins2007} that results in a median obscuration of $f^{\mathrm{AGN}}_{\mathrm{esc}}\sim 0.19_{-0.08}^{+0.10}$ for all AGN in TNG50 at $z=3$. The sub- and superscript denote the interval of the central 99.73\% of AGN. Higher obscuration factors typically occur for AGN in higher mass halos.

For AGN with low accretion $\dot{m} / \dot{m}_\mathrm{Eddington}<0.002$, the radiative feedback is set to zero in the TNG model. The radiative output of AGN differs substantially with redshift. While at $z=0.0$ the vast majority of halos are inactive given this threshold, only $\lesssim 5\%$ are radiatively inactive for the studied redshift range $z\geq 2$.

More details of the implemented radiative feedback model in TNG can be found in~\cite{vogelsberger13,weinberger17,weinberger18}.

TNG incorporates "on-the-fly" ionizing radiation from SMBH in each time step as given in Equation~\eqref{eq:L_UV_AGN}, which impacts gas cooling, temperature, and ionization state. For the gas cooling, these local sources' photoionization and photoheating are incorporated. The model assumes the bolometric intensity as $J^\mathrm{SMBH}_\mathrm{UV,esc}\propto L^\mathrm{SMBH}_\mathrm{UV,esc}/r^2$ at a distance $r$ of each nearby AGN under the assumption that the gas is optically thin to SMBH radiation.
Photoionization and photoheating are calculated as a superposition of the local ionizing radiation from AGN and the uniform metagalactic background. For halos with a radiating black hole in TNG, the AGN field dominates over the UV background within the halo.

We will now assess the impact of the SMBHs' ionizing radiation on the halos' surrounding gas and the modifications to their \Lya halos. To do this, we split the sample in halos with and without SMBHs. In TNG50, this is equal to splitting the samples by AGN radiative feedback activity.

In Figure~\ref{fig:rprofiles_AGN_hydrostate} we show the median radial profiles for the neutral hydrogen fraction $x_\mathrm{HI}$ and temperature $T$ for the sample of halos in the stellar mass range \smassmini at $z=3.0$ in TNG50. For this mass range, we have a roughly equal sample of halos with (148) and without (122) SMBH activity (also see cutoff in Figure~\ref{fig:L_AGN_vs_L_SF}). In addition to the central 68 percentiles as shaded area, we also show the 2 median arcsecond radius given the virial radii, along the x-axis (vertical dotted line).

Note that only 1 of the 148 AGN is radiatively inactive. Hence, for our sample, AGN presence nearly always implies a significant ionizing flux in the surrounding CGM according to Equation~\eqref{eq:L_UV_AGN}. As a consequency of the SMBH (radiative) feedback, we find a strongly suppressed neutral hydrogen content in affected halos. At the same time, the temperature is significantly higher outside the star-forming regions at $r>0.2 r_\mathrm{vir}$. Given the decreased neutral hydrogen fraction, we expect a boosted \Lya emission from recombinations, and in particular a boosted \Lya emission from collisional excitations given the strong temperature dependency of the latter.

\begin{figure}
\centering
  \includegraphics[width=1.0\linewidth]{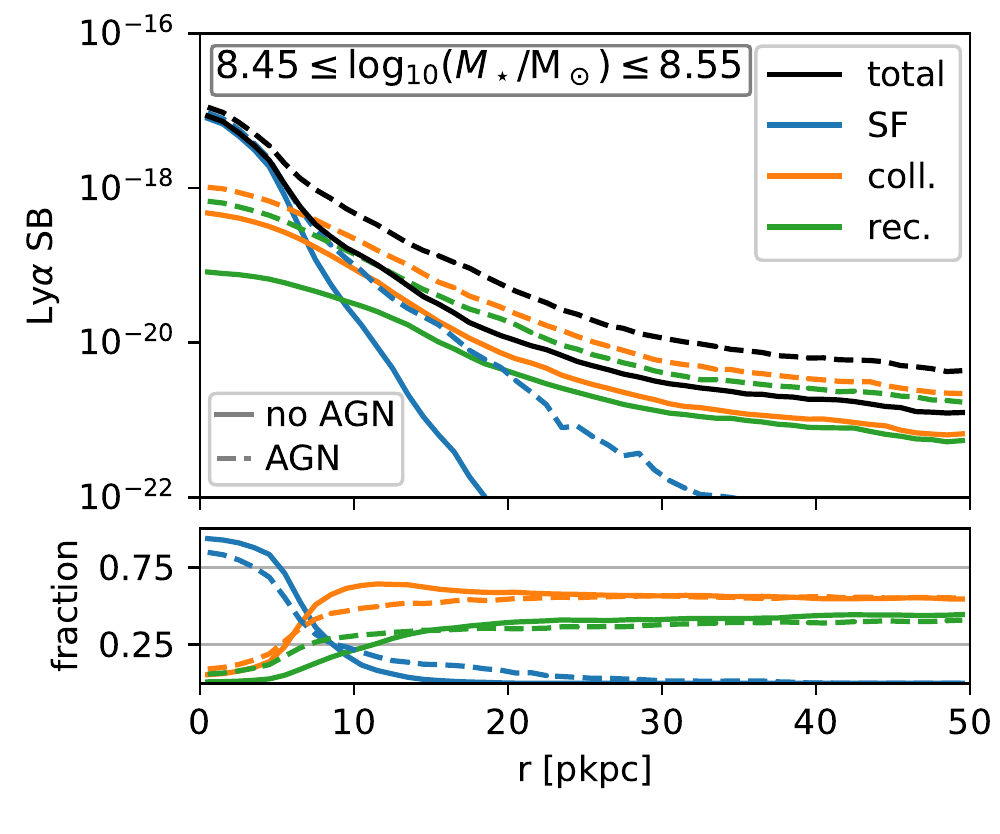}
\caption{Median stacked Ly$\alpha$ surface brightness radial profiles from intrinsic emission (i.e. no scattering) at $z=3$ for halos with \smassdfltlow{} decomposed into different emission sources (upper panel), and the relative fraction of each (lower panel) in TNG50. Dashed lines show the halos with AGN activity (148 halos), while solid lines show the halos without AGN activity (122 halos). Without scattering, collisional excitation dominates the radial profiles outside of the star-forming regions $r_0\gtrsim5\ $pkpc irrespective of AGN activity. AGN activity however largely boosts emission from collisional excitations and recombinations.}
\label{fig:rprofiles_AGN_int}
\end{figure}

In Figure~\ref{fig:rprofiles_AGN_int} we show the \textit{intrinsic} median \Lya surface brightness radial profiles for $z=3.0$ and \smassmini split into their respective emission mechanisms and divided into the sample with and without SMBH activity similar to the dashed lines in Figure~\ref{fig:rprofiles_mechanisms}. 
Without scattering, we generally find collisional excitations to be the dominant emission mechanisms for the majority of halos above $7\ $pkpc irrespective of SMBH activity. Particularly recombinations are strongly boosted in the presence of a SMBH for $r_0\lesssim 20\ $pkpc. Interestingly, while the overall emission from collisional excitations is boosted, the relative fraction decreases not just because of the higher recombination rates but also due to a larger fraction of emission from star-forming regions. 

To understand latter point, we stress that the underlying samples differ. Particularly, the sample hosting SMBHs, even though with a similar stellar mass, are typically more massive with a median total halo mass $\log_{10} \left(M_h/\rm{M}_\odot\right)$ higher by about $0.2\ $dex and a significant scatter towards higher masses. The high mass objects show a larger amount of satellites contributing to the additional emission from star-formation in the halos' outskirts.

\begin{figure}
\centering
  \includegraphics[width=1.0\linewidth]{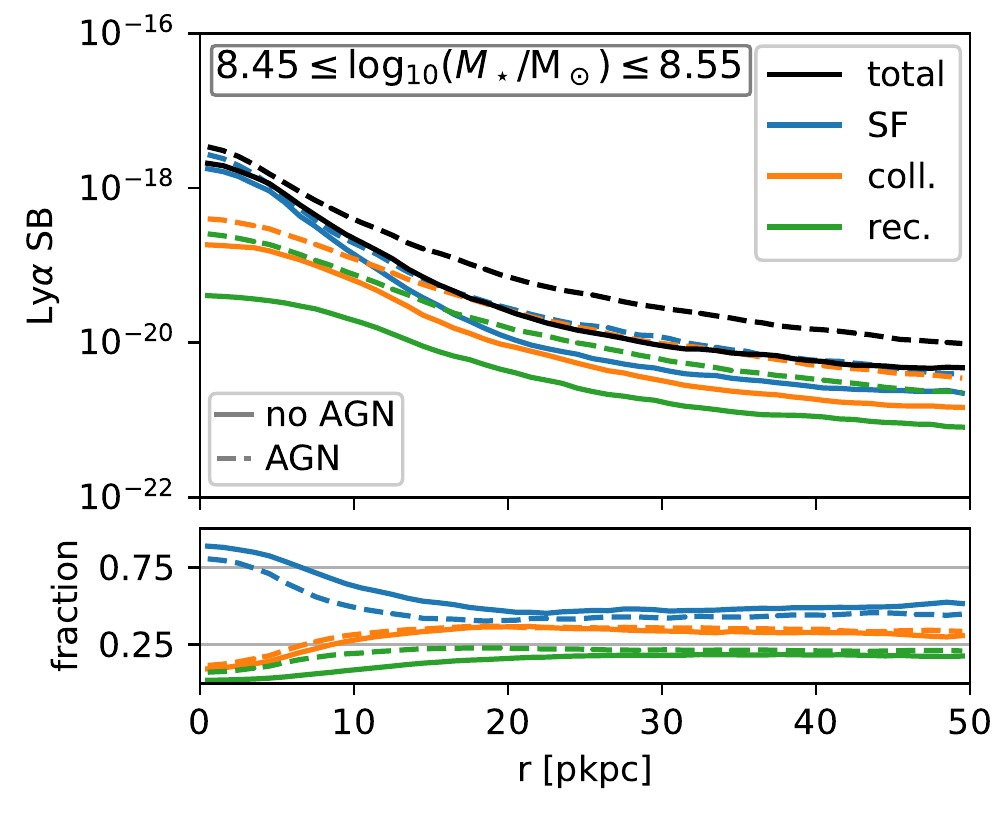}
\caption{Same as Figure~\ref{fig:rprofiles_AGN_int}, but for the processed emission (i.e. with scatterings). We observe a similar behaviour for both samples as in Figure~\ref{fig:rprofiles_mechanisms}. Slight differences in shape seem to arise from boosted collisional excitations and recombinations for $r_0<20\ $pkpc relative to the contribution from star-forming cells, leading to a slightly flatter slope. Star-formation remains the dominant contributor to the radial profiles across the shown radii.}
\label{fig:rprofiles_AGN}
\end{figure}

Figure~\ref{fig:rprofiles_AGN} shows the \textit{processed} median radial profiles, i.e. after running the \Lya radiative transfer for the intrinsic emission shown in~\ref{fig:rprofiles_AGN_int}. While all emission mechanisms remain boosted in absolute terms when SMBH activity is present, collisional excitations and particularly recombinations become more important in relative terms. However, irrespective of SMBH activity the radial profiles remain dominantly sourced by emission from star-forming cells. At $r_0\gtrsim 20\ $pkpc, contributions from collisional excitations become close to equally important to photons from star-forming regions. However, this trend is the same for both the AGN and no-AGN sample.

From the median profiles it appears that there is a slight additional flattening for the AGN hosting subsample compared to the no-AGN sample. To quantify the difference of the LAHs' shape, we calculate the individual exponential scale lengths $r_0$ and find an overall increase of roughly $11\%$ from $3.6_{-0.9}^{+1.2}\ $pkpc (no AGN) to $4.0_{-1.0}^{+1.4}\ $pkpc (AGN).

This increase is however not solely driven by the ionizing AGN radiation but also by the different halo population, given the fixed stellar mass ranges of the sample. If we instead constrain the total halo mass to be within a fixed range of \hmassmini, which corresponds to the central $\sim 68\%$ of the non-AGN \smassmini sample, we obtain a reversed trend with $r_0$ being $3.8_{-0.9}^{+1.4}\ $pkpc ($3.6_{-0.8}^{+1.7}\ $pkpc) for the sample without (with) AGN.

In conclusion, the impact of ionizing radiation from SMBHs in TNG appears to be significant in terms of ionization state, temperature and thus intrinsic \Lya emission of the gas, but the findings on the LAH shape, particularly through the scale radius $r_0$, remain unchanged irrespective of modelled SMBH activity. We therefore do not split our sample within the main body with respect to the presence of AGN.

Note that here, we only discuss the AGN's impact through its ionizing budget onto the surrounding. In particular, we do not discuss nor implement a description for the unresolved \Lya emission from AGN itself at this point. We would expect that this emission would scatter outwards and produce a contribution to the LAH of similar shape to that of the star-formation, given the concentrated emission source scattering into the surrounding CGM. As differences in the intrinsic emission in terms of spatial and spectral distribution exist, this would need to be explored in future investigations.

\subsubsection{Stellar Populations}
\label{sec:stars}

TNG does not incorporate the ionizing flux from local stellar populations as is done for SMBHs. Hence, the possible impact of those sources on our predicted LAH profiles cannot be assessed within the existing simulations. However, we can derive an estimate of the upper limit of this effect, by comparing to the local ionization from AGN.

The ionizing luminosity escaping from the stellar populations, adopting a formulation consistent with our model, is given as:
\begin{align}
\label{eq:L_UV_SF}
    L^{\mathrm{SF}}_\mathrm{UV,esc} = \frac{f_\mathrm{UV,esc}}{f_\mathrm{B} \left(1-f_\mathrm{UV,esc}\right)} \frac{\langle E_{\gamma,\mathrm{UV}}\rangle}{E_{\gamma,\Lya}}\  \epsilon_\mathrm{SF}\ V_\star
\end{align}

As in Equation~\eqref{eq:lumSF}, we assume no dust and \textit{Case B} recombination with $f_\mathrm{B}=0.68$ being the conversion factor from ionizing to \Lya photons.
$\langle E_{\gamma,\mathrm{UV}}\rangle/E_{\gamma,\Lya}$ is the ratio of the average ionizing photon energy in the population and the \Lya line transition energy. We assume $f_\mathrm{UV,esc}=0.1$ in Equation~(17) of \cite{Dijkstra2019}. 

\begin{figure}
\centering
  \includegraphics[width=1.0\linewidth]{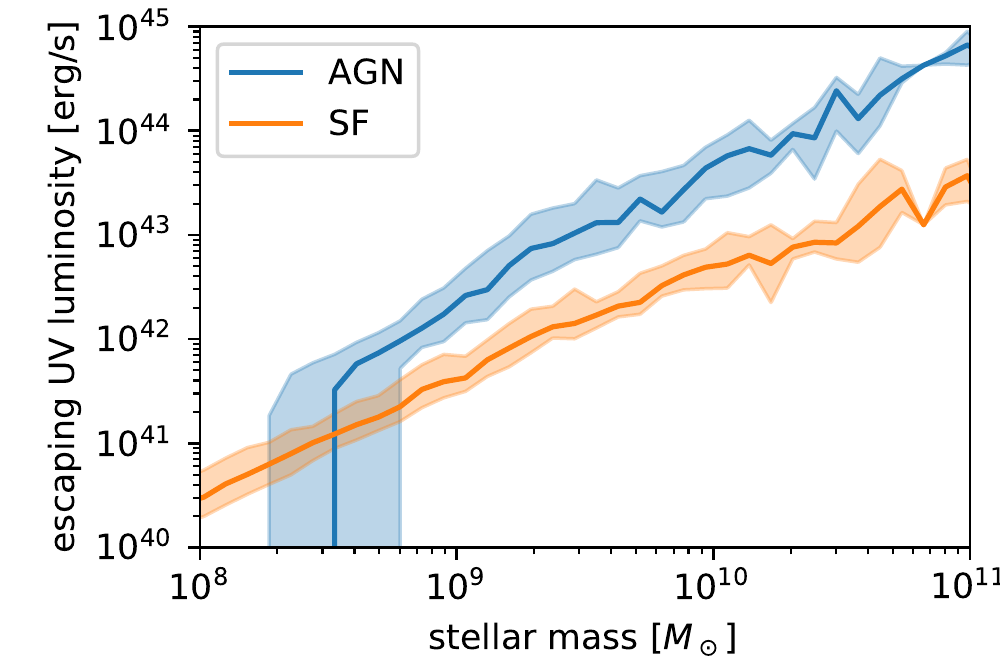}
  \caption{We show the median of the UV luminosity escaping the galaxy at a given stellar mass of the host halo at $z=3.0$ in TNG50. The UV luminosities for star-formation (SF) and SMBH are derived from Equations~\eqref{eq:L_UV_AGN}/\eqref{eq:L_UV_SF} respectively.}
\label{fig:L_AGN_vs_L_SF}
\end{figure}

In Figure~\ref{fig:L_AGN_vs_L_SF} we show the escaping UV radiation from stellar sources and AGN as a function of the stellar mass of the host halo at $z=3.0$ in TNG50. We find AGN activity for the majority of halos with a stellar mass above $\log_{10} \left(M_\star/\rm{M}_\odot\right) \gtrsim 8.5$.
Above this mass, we find an approximate power law $L_\mathrm{UV,esc} = L_0 \cdot \left(\frac{M_\star}{10^{8.5}\mathrm{M}_\odot}\right)^\alpha$ with $L_0=10^{41.7}$ and $\alpha=1.31$ ($L_0=10^{41.1}$, $\alpha=0.99$) for AGN (SF).
We thus find that $L^\mathrm{AGN}_\mathrm{UV,esc}\gtrsim 6\cdot L^\mathrm{SF}_\mathrm{UV,esc}$ at $z=3.0$ for the bulk of halos with AGN activity and a growing disparity between the luminosities with larger mass given the larger slope for AGN. The discrepancy grows (shrinks) at lower (higher) redshifts across the stellar mass range.

Changing $f_\mathrm{UV,esc}$ or adopting different assumptions concerning e.g. metallicity, binary fraction or IMF can boost the stellar populations' escaping UV luminosity. Given the large margin at $z=3.0$ between AGN and stellar luminosities, qualitative findings should be robust. However, such adjustments might lower the redshift at which stellar populations' UV luminosity becomes dominant into the upper studied redshift range.  
For the given UV luminosities, the AGN ionizing flux integrated over the halo population dominates over the ionizing flux from stellar populations. This ratio peaks at $z=2$ in TNG50, where AGN are most relevant, while stellar populations start to dominate the overall ionizing flux at higher redshifts $z \gtrsim 6$.

We found that the ionizing flux of SMBHs, if present, dominates over the ionizing flux from stellar populations in the same host halo. A consideration of the impact photoionization and photoheating thus needs to primarily address the impact of AGN, for which a simplified description is indeed implemented in TNG50.

In our fiducial sample with stellar masses of \smassdflt{}, 87\% of halos host at least one SMBH with a radiation field according to Eqn.~\ref{eq:L_UV_AGN}. Therefore, our analysis contains the primary ionization source. As demonstrated in Appendix~\ref{sec:AGN}, while the gas becomes significantly hotter and ionized, the LAH size measurements and our conclusions remain largely unaffected. As a consequence, we also find the inclusion of stellar ionizing sources, with overall weaker UV luminosities, would have a small impact on our overall findings.

\subsection{Impact of a Wavelength Window}
\label{sec:spectra}

\begin{figure}
\centering
  \includegraphics[width=1.0\linewidth]{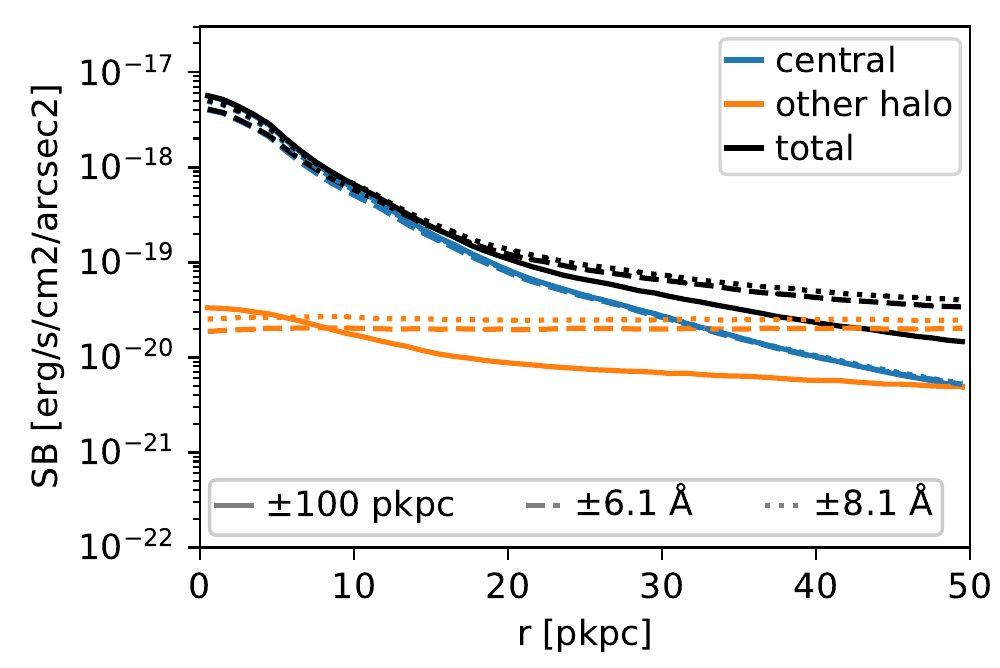}
\caption{Median stacked radial Ly$\alpha$ profile for galaxies with stellar masses \smassdflt{} at $z=3$ in TNG50. We decompose the radial profiles into their two dominant origins from central galaxies (blue) and other halos (orange). For the solid line, we show the fiducial method throughout this paper that ignores spectral information and integrates all photon contributions that scatter last in the $\pm 100\ $pkpc around the halo's position. The dashed and dotted lines show the contributions when incorporating spectral information in a $\sim 12\ \angstrom$ (dashed) and $\sim 16\ \angstrom$ (dotted) observed wavelength window around the \Lya line center. The central surface brightness is slightly suppressed when considering spectral information due to some emitters with a spectral diffusion from the central galaxies exceeding the imposed wavelength window. At large radii, the wavelength windows lead to a stronger flattening due to the larger physical depth.}
\label{fig:rprofiles_redshiftspace}
\end{figure}

For simplicity, we ignored spectral information throughout this paper and instead sum all photons scattering last within $\pm 100\ $pkpc depth around the halos' center. Here, we show the difference when instead using a simple spectral bandwidth from the photons' spectral information incorporating Hubble flow, peculiar velocity and spectral diffusion. Just as in Appendix~\ref{sec:ionizingsources}, we consider contributions from all three implemented emission sources and do not neglect the rescattered light from other halos as done in the other appendices.
We note that a fair comparison will need to carefully match the various methodologies used in observational studies. For example~\cite{leclercq17} uses an adaptive spectral bandwidth per emitter.

In Figure~\ref{fig:rprofiles_redshiftspace} we show the median stacked radial profiles using the fiducial depth integration based on the last scattering of photons (solid line) opposed to a fixed spectral bandwidth of $\pm 6.1\ \angstrom$ (dashed line) and $\pm 8.1\ \angstrom$ (solid line) for the overall radial profile and split into the dominant emission origins (central galaxies in blue, other halos in orange). Ignoring spectral diffusion, we effectively model an adaptive spectral bandwidth to capture potentially large spectral diffusion. Thus, a fixed spectral bandwidth leads to a slightly suppressed median radial profile in Figure~\ref{fig:rprofiles_redshiftspace}.

The scale lengths $r_0=4.2_{-1.1}^{+2.0}\ $pkpc ($\pm 6.1\ \angstrom$) and $4.0_{-1.0}^{+1.9}\ $pkpc ($\pm 8.1\ \angstrom$) for the fixed bandwidth window slightly increase compared to the fiducial setup due to the exclusion of \Lya contributions that diffused outside of these windows. 

In Figure~\ref{fig:rprofiles_origins} and~\ref{fig:veryextended} we demonstrate a flattening of the radial profiles that is dominated by contributions from other halos. This flattening will therefore be heavily influenced by the environment in a chosen field of view and the methodology for its detection and stacking. A fair comparison will thus require a thorough reproduction of factors such as field overdensity, source masking and chosen wavelength depth along the line of sight. Generally, a fixed bandwidth as in Figure~\ref{fig:rprofiles_redshiftspace} causes a larger flattening due to the larger relative background contribution from other halos which is expected to scale roughly linear with the integration depth. The integration depth from the differential Hubble flow corresponds to $2.4\ $pMpc and $3.2\ $pMpc respectively compared with the $0.2\ $pMpc in the fiducial setup. Future observational comparisons will benefit from close attention to the chosen wavelength depth along with a careful account of the background subtraction.

\label{lastpage}
\end{document}